\documentclass{siamltex}
\usepackage{amsmath, amssymb}
\usepackage{color, epsfig}
\usepackage{verbatim}
\usepackage{ifpdf}

\title{Logarithmic Lower Bounds in the Cell-Probe Model%
\thanks{This work is based on two conference publications by the same
  authors: ``Lower Bounds for Dynamic Connectivity'', appearing in
  Proc.\ 36th ACM Symposium on Theory of Computing (STOC'04),
  pp.~546--553, and ``Tight Bounds for the Partial-Sums Problem'',
  appearing in Proc.\ 15th ACM-SIAM Symposium on Discrete Algorithms
  (SODA'04), pp.~20--29.}}

\author{Mihai P\v{a}tra\c{s}cu%
  \thanks{MIT Computer Science and Artificial Intelligence Laboratory,
      32 Vassar St., Cambridge, MA 02139, USA,
      \{\texttt{mip},~\texttt{edemaine}\}\texttt{@mit.edu}}%
\and
Erik D. Demaine%
  \footnotemark[2]}

\newcommand{\func}[1] {\textnormal{\tt\scshape#1}}
\newcommand{\poly} {\mathrm{poly}}
\newcommand{\eps} {\varepsilon}
\newcommand{\twodots} {\mathinner{\ldotp\ldotp}}

\def\rep{\mathop{\rm rep}\nolimits}
\def\len{\mathop{\rm len}\nolimits}

\let\latexcite=\cite
\def\cite{\nolinebreak\latexcite}
\let\latexref=\ref
\def\ref{\nolinebreak\latexref}

\usepackage
  [breaklinks,bookmarks,bookmarksnumbered,bookmarksopen,bookmarksopenlevel=2]
  {hyperref}
{\makeatletter \hypersetup{pdftitle={\@title}}}

{\makeatletter
 \gdef\xxxmark{%
   \expandafter\ifx\csname @mpargs\endcsname\relax 
     \expandafter\ifx\csname @captype\endcsname\relax 
       \marginpar{xxx}
     \else
       xxx 
     \fi
   \else
     xxx 
   \fi}
 \gdef\xxx{\@ifnextchar[\xxx@lab\xxx@nolab}
 \long\gdef\xxx@lab[#1]#2{{\bf [\xxxmark #2 ---{\sc #1}]}}
 \long\gdef\xxx@nolab#1{{\bf [\xxxmark #1]}}
}

\renewcommand{\th}{\ifmmode{^{\textrm{th}}}\else{\textsuperscript{th}\ }\fi}

\begin{document}

\maketitle

\begin{abstract}
  We develop a new technique for proving cell-probe lower bounds on
  dynamic data structures.  This technique enables us to prove an
  amortized randomized $\Omega(\lg n)$ lower bound per operation for
  several data structural problems on $n$ elements, including partial
  sums, dynamic connectivity among disjoint paths (or a forest or a
  graph), and several other dynamic graph problems (by simple
  reductions).  Such a lower bound breaks a long-standing barrier of
  $\Omega(\lg n / \lg\lg n)$ for any dynamic language membership
  problem.  It also establishes the optimality of several existing
  data structures, such as Sleator and Tarjan's dynamic trees.  We
  also prove the first $\Omega(\log_B n)$ lower bound in the
  external-memory model without assumptions on the data structure
  (such as the comparison model).  Our lower bounds also give a
  query-update trade-off curve matched, e.g., by several data
  structures for dynamic connectivity in graphs.  We also prove
  matching upper and lower bounds for partial sums when parameterized
  by the word size and the maximum additive change in an update.
\end{abstract}


\begin{AM}
68Q17
\end{AM}

\pagestyle{myheadings}
\thispagestyle{plain}
\markboth{M. P\v{A}TRA\c{S}CU AND E. D. DEMAINE}{LOGARITHMIC LOWER BOUNDS}

\section{Introduction}

The cell-probe model is perhaps the strongest model of computation for
data structures, subsuming in particular the common word-RAM model.
We suppose that the memory is divided into fixed-size cells (words),
and the cost of an operation is just the number of cells it reads or
writes.  Typically we think of the cell size as being around $\lg n$
bits long, so that a single cell can address all $n$ elements in the
data structure.  (Refer to Section \ref{models} for a precise
definition of the model.)  While unrealistic as a model of computation
for actual data structures, the generality of the cell-probe model
makes it an important model for lower bounds on data structures.


Previous cell-probe lower bounds for data structures fall into two
categories of approaches.  The first approach is based on
communication complexity. Lower bounds for the predecessor problem
\cite{ajtai88predecessor, miltersen99asymmetric, beame02predecessor,
sen-roundelim} are perhaps the most successful application of this
idea. Unfortunately, this approach can only be applied to problems
that are hard even in the static case. It also requires queries to
receive a parameter of $\omega(\lg n)$ bits, which is usually
interpreted as requiring cells to have $\omega(\lg n)$ bits.
For problems that are hard only in the dynamic case, all lower bounds
have used some variation of the chronogram method of Fredman and Saks
\cite{fredman89cellprobe}.  By design, this method cannot prove a
trade-off between the query time $t_q$ and the update time $t_u$
better than $t_q \lg t_u = \Omega(\lg n)$, which was achieved for the
marked-ancestor problem (and consequently many other problems) in
\cite{alstrup98marked}.  This limitation on trade-off lower bounds
translates into an $\Omega(\lg n / \lg\lg n)$ limitation on lower
bounds for both queries and updates provable by this technique.  The
$\Omega(\lg n / \lg\lg n)$ barrier has been recognized as an important
limitation in the study of data structures, and was proposed as a
major challenge for future research in a recent survey
\cite{miltersen99survey}.

This paper introduces a new technique for proving cell-probe lower
bounds on dynamic data structures.  With this technique we establish
an $\Omega(\lg n)$ lower bound for either queries or updates in
several natural and well-studied problems, in particular, maintaining
partial (prefix) sums in an array and dynamic connectivity among
disjoint paths (or a forest or a graph).  (We detail the exact
problems we consider, and all results we obtain, in Section
\ref{results}; we summarize relevant previous results on these
problems in Section \ref{history}.) These lower bounds establish the
optimality of several data structures, including the folklore $O(\lg
n)$ balanced tree data structure for partial sums, and Sleator and
Tarjan's dynamic trees data structure (which in particular maintains
dynamic connectivity in a forest).

We also prove a trade-off lower bound of $t_q \lg \frac{t_u}{t_q} =
\Omega(\lg n)$.%
\footnote{Throughout this paper, $\lg x$ denotes $\log_2 (2+x)$,
  which is positive for all $x \geq 0$
  (an important property in this bound and several others).}
This trade-off turns out to be the right answer for
our problems, and implies the $\Omega(\lg n)$ bound on the worst of
queries and updates.  In addition, we can prove a symmetric trade-off
$t_u \lg \frac{t_q}{t_u} = \Omega(\lg n)$.  As mentioned above, it is
fundamentally impossible to achieve such a trade-off using the
previous techniques.

We also refine our analysis of the partial-sums problem beyond just
the dependence on~$n$.  Specifically, we parameterize by $n$, the
number $b$ of bits in a word, and the number $\delta$ of bits in an
update.  Naturally, $\delta \leq b$, but in some applications,
$\delta$ is much smaller than~$b$.  We prove tight upper and lower
bounds of $\Theta(\frac{\lg n}{\lg(b/\delta)})$ on the worst of
queries and updates.  This result requires improvements in both the
upper bounds and the lower bounds.  In addition, we give a tight
query/update trade-off: $t_q \left( \lg \frac{b}{\delta} + \lg
\frac{t_u}{t_q} \right) = \Theta(\lg n)$.  The tightness of this
characterization is particularly unusual given its dependence on five
variables.

The main idea behind our lower-bound technique is to organize time
(the sequence of operations performed on the data structure) into a
complete tree.  The heart of the analysis is an encoding/decoding
argument that bounds the amount of information transferred between
disjoint subtrees of the tree: if few cell are read and written, then
little information can be transferred.  The nature of the problems of
interest requires at least a certain amount of information transfer
from updates to queries, providing a lower bound on the number of
cells read and written.  This main idea is developed first in Section
\ref{simple-sums} in the context of the partial-sums problem, where we
obtain a short (approximately three-page) proof of an $\Omega(\lg n)$
lower bound for partial sums. Compared to the lower bounds based on
previous techniques, our technique leads to relatively clean proofs
with minimal combinatorial calculation.

We generalize this basic approach in several directions to obtain our
further lower bounds.  In Section \ref{verify-sums}, we show how our
technique can be extended to handle queries with binary answers (such
as dynamic connectivity) instead of word-length answers (such as
partial sums).  In particular, we obtain an $\Omega(\lg n)$ lower
bound for dynamic connectivity in disjoint paths.  We also show how to
use our lower-bound technique in the presence of nondeterminism or
Monte Carlo randomization.  In Section \ref{highb}, we show how our
technique can be further extended to handle updates asymptotically
smaller than the word size, in particular obtaining lower bounds for
the partial-sums problem when $\delta < b$ and for dynamic
connectivity in the external-memory model.  This last section develops
the most complicated form of our technique.

The final few sections contain complementary results to this main flow
of the lower-bound technique.  In Section \ref{upperbnds}, we give
tight upper bounds for the partial-sums problem.  The data structure
is based on a few interesting ideas that enable us to eliminate the
precomputed tables from previous approaches.  In Section
\ref{reductions}, we prove some easy reductions from dynamic
connectivity to other dynamic graph problems, transferring our lower
bounds to these problems.  Finally, we conclude in Section
\ref{theend} with a list of open problems.

\section{Results} \label{results}

In this section we give precise descriptions of the problems we
consider, our results, and a brief synopsis of how these results
compare to previous work.  (Section \ref{history} gives a more
detailed historical account.)

\subsection{The Partial-Sums Problem}

This problem asks to maintain an array $A[1 \twodots n]$ of $n$
integers subject to the following operations:

\smallskip

\begin{description} 
\item[\quad $\func{update}(k, \Delta)$:] modify $A[k] \gets \Delta$.

\item[\quad $\func{sum}(k)$:] returns the partial sum $\sum_{i=1}^{k}
  A[i]$.

\item[\quad $\func{select}(\sigma)$:] returns an index $i$ satisfying
  $\textrm{sum}(i - 1) < \sigma \leq \textrm{sum}(i)$. To guarantee
  uniqueness of the answer, we require that $A[i] > 0$ for all~$i$.
\end{description}

\smallskip

Besides $n$, the problem has several interesting parameters.  One
parameter is $b$, the number of bits in a cell (word).  We assume that
every array element and sum fits in a cell.  Also, we assume that $b =
\Omega(\lg n)$.  Another parameter is $\delta$, the number of bits
needed to represent an argument $\Delta$ to \func{update}. Naturally,
$\delta$ is bounded above by $b$; however, it is traditional (see,
e.g.~\cite{raman01succinct}) to consider a separate parameter $\delta$
because it is smaller in many applications. We write $t_u$ for the
running time of \func{update}, $t_q$ for the running time of
\func{sum}, and $t_s$ for the running time of \func{select}.

We first study the unrestricted case when $\delta = \Omega(b)$:

\begin{theorem} \label{sums-thm}
Consider any cell-probe data structure for the partial-sums problem
that may use amortization and Las Vegas randomization.  If $\delta =
\Omega(b)$, then the following trade-offs hold:

\vspace{-0.7cm}
\begin{eqnarray*}
t_q \lg( t_u / t_q ) = \Omega(\lg n); & \qquad & 
  t_u \lg( t_q / t_u ) = \Omega(\lg n);
\\
t_s \lg( t_u / t_s ) = \Omega(\lg n); & \qquad & 
  t_u \lg( t_s / t_u ) = \Omega(\lg n).
\end{eqnarray*}

\end{theorem}

The trade-off curves are identical for the \func{select} and
\func{sum} operations.  The first branch of each trade-off is relevant
when queries are faster than updates, while the second branch is
relevant when updates are faster.  The trade-offs imply the
long-sought logarithmic bound for the partial-sums problem: $\max\{t_u,
t_q\} = \Omega(\lg n)$.  The best previous bound, by Fredman and Saks
\cite{fredman89cellprobe}, was $t_q \lg(b t_u) = \Omega(\lg n)$,
implying $\max\{t_u, t_q\} = \Omega(\lg n / \lg b)$.  The trade-off
curves between $t_u$ and $t_q$ also hold in the group model of
computation, where elements of the array come from a black-box group
and time is measured as the number of algebraic operations.  The best
previous bound for this model was $\Omega(\lg n / \lg\lg n)$, also by
Fredman and Saks \cite{fredman89cellprobe}.

A classic result achieves $t_u = t_q = t_s = O(\lg n)$.  For the
\func{sum} query, our entire trade-off curve can be matched (again,
this is folklore; see the next section on previous work).  For
\func{select}, the trade-offs cannot be tight for the entire range of
parameters, because even for polynomial update times, there is a
superconstant lower bound on the query time for the predecessor
problem \cite{ajtai88predecessor,sen-roundelim}.

We also analyze the case $\delta = o(b)$.  We first give the following
lower bounds:

\begin{theorem} \label{highb-thm}
Consider any cell-probe data structure for the partial-sums problem
using amortization and Las Vegas randomization.  The following
trade-offs hold: $t_q \left( \lg \frac{b}{\delta} + \lg
\frac{t_u}{t_q} \right) = \Omega(\lg n)$ and $t_s \left( \lg
\frac{b}{\delta} + \lg \frac{t_u}{t_s} \right) = \Omega(\lg n)$
\end{theorem}

In this case, we cannot prove a reverse trade-off (when updates are
faster than queries).  This trade-off implies the rather interesting
lower bound $\max\{t_u, t_q\} = \Omega( \frac{\lg n}{\lg(b/\delta)})$,
and similarly for $t_s$. When $\delta = \Theta(b)$, this gives the
same $\Omega(\lg n)$ as before. We give new matching upper bounds:

\begin{theorem}
There exists a data structure for the partial-sums problem achieving
$t_u = t_q = t_s = O\left( \frac{\lg n}{\lg(b / \delta)} \right)$. The
data structure runs on a Random Access Machine, is deterministic, and
achieves worst-case bounds.
\end{theorem}

Our upper bounds can handle a slightly harder version of the problem,
where $\func{update}(i, \Delta)$ has the effect $A[i] \gets A[i] +
\Delta$. Thus, we are not restricting each $A[i]$ to $\delta$ bits,
but just mandate that they don't grow by more than a $\delta$-bit term
at a time.  Several previous results \cite{dietz89sums,
raman01succinct, hon03sums} achieved $O(\lg n / \lg\lg n)$ bounds for
$\delta = O(\lg\lg n)$.  None of these solutions scale well with
$\delta$ or~$b$, because they require large precomputed tables.

This result matches not only the previous lower bound on the hardest
operation, but actually helps match the entire trade-off of Theorem
\ref{highb-thm}. Indeed, the trade-off lower bound shows that there is
effectively no interesting trade-off when $\delta = o(b)$: when
$\frac{b}{\delta} \ge \frac{t_u}{t_q}$, the tight bound is $t_q =
\Theta(\frac{\lg n}{\lg (b/\delta)})$, matched by our structure; when
$\frac{b}{\delta} < \frac{t_u}{t_q}$, the tight bound is $t_q =
\Theta(\frac{\lg n}{\lg (t_u / t_q)})$, matched by the classic result
which does not depend on~$\delta$.  Thus, we obtain an unusually precise
understanding of the problem, having a trade-off that is tight in all
five parameters.

\subsection{Dynamic Connectivity}

This problem asks to maintain an undirected graph with a fixed set of
$n$ vertices subject to the following operations:

\smallskip

\begin{description} 
\item[\quad $\func{insert}(u, v)$:] insert an edge $(u,v)$ into the
  graph.

\item[\quad $\func{delete}(u, v)$:] delete the edge $(u,v)$ from the
  graph.

\item[\quad $\func{connected}(u, v)$:] test whether $u$ and $v$ lie in
  the same connected component.
\end{description}

\smallskip

We write $t_q$ for the running time of \func{connected}, and $t_u$ for the
running time of \func{insert} and \func{delete}.
It makes the most sense to study this problem in the cell-probe model
with $O(\lg n)$ bits per cell, because every quantity in the problem
occupies $O(\lg n)$ bits.

We prove the following lower bound:

\begin{theorem}
Any cell-probe data structure for dynamic connectivity satisfies the
following trade-offs: $t_q \lg(t_u / t_q) = \Omega(\lg n)$ and
$t_u \lg(t_q / t_u) = \Omega(\lg n)$.
These bounds hold under amortization, nondeterministic queries, and Las Vegas
randomization, or under Monte Carlo randomization with error probability
$n^{-\Omega(1)}$.
These bounds hold even if the graph is always a disjoint union of paths.
\end{theorem}

This lower bound holds under very broad assumptions. It allows for
nondeterministic computation or Monte Carlo randomization (with
polynomially small error), and holds even for paths (and thus for
trees, plane graphs etc.). The trade-offs we obtain are identical to
the partial-sums problem.  In particular, we obtain that $\max\{t_u,
t_q\} = \Omega(\lg n)$.

An upper bound of $O(\lg n)$ for trees is given by the famous dynamic
trees data structure of Sleator and Tarjan \cite{sleator83dynamic}.
In addition, the entire trade-off curve for $t_u = \Omega(t_q)$ can be
matched for trees.  For general graphs, Thorup \cite{thorup00connect} gave
an almost-matching upper bound of $O(\lg n (\lg\lg n)^3)$.  For any
$t_u = \Omega(\lg n (\lg\lg n)^3)$, his data structure can match our
trade-off.

Dynamic connectivity is perhaps the most fundamental dynamic graph problem.
It is relatively easy to show by reductions that our bounds hold
for several other dynamic graph problems.
Section \ref{reductions} describes such reductions for deciding connectivity
of the entire graph, minimum spanning forest, and planarity testing. 
Many data structural problems on undirected graphs have
polylogarithmic solutions, so our bound is arguably interesting for these
problems.  Some problems have logarithmic solutions for special cases
(such as plane graphs), and our results prove optimality of those data
structures.

We also consider dynamic connectivity in the external-memory model.
Let $B$ be the page (block) size, i.e., the number of $(\lg n)$-bit cells
that fit in one page.  We prove the following lower bound:

\begin{theorem}
A data structure for dynamic connectivity in the external-memory model
with page size $B$ must satisfy $t_q \left( \lg B + \lg
\frac{t_u}{t_q} \right) = \Omega(\lg n)$.  This bound allows for
amortization, Las Vegas randomization, and nondeterminism, and holds
even if the graph is always a disjoint union of paths.
\end{theorem}

Thus we obtain a bound of $\max\{t_u, t_q\} = \Omega(\log_B n)$.
Although bounds of this magnitude are ubiquitous in the
external-memory model, our lower bound is the first that holds in a
general model of computation, i.e., allowing data items to be
manipulated arbitrarily and just counting the number of page
transfers.  Previous lower bounds have assumed the comparison model or
indivisibility of data items.

It is possible to achieve an $O(\log_B n)$ upper bound for a forest,
by combining Euler tour trees with buffer trees \cite{arge03buffer}.
As with the partial-sums problem, this result implies that our entire
trade-off is tight for trees: for $B \ge t_u / t_q$, this solution is
optimal; if the term in $t_u / t_q$ dominates, we use the classic trade-off,
which foregoes the benefit of memory pages.

\section{Previous Work} \label{history}

In this section we detail the relevant history of cell-probe lower bounds
in general and the specific problems we consider.

\subsection{Cell-Probe Lower Bounds}

Fredman and Saks \cite{fredman89cellprobe} were the first to prove
cell-probe lower bounds for dynamic data structures. They developed
the chronogram technique, and used it to prove a lower bound of
$\Omega(\lg n / \lg b)$ for the partial-sums problem in $\mathbb{Z} /
2\mathbb{Z}$ (integers modulo~$2$, where elements are bits and
addition is equivalent to binary exclusive-or).  This bound assumes $b
\geq \lg n$ so that an index into the $n$-element array fits in a
word; for the typical case of $b = \Theta(\lg n)$, it implies an
$\Omega(\lg n / \lg \lg n)$ lower bound.  Fredman and Saks also obtain
a trade-off of $t_q = \Omega(\frac{\lg n}{\lg b + \lg t_u})$.

There has been considerable exploration of what the chronogram
technique can offer.  Ben-Amram and Galil \cite{benamram01sums}
reprove the lower bounds of Fredman and Saks in a more formalized
framework, centered around the concepts of problem and output
variability.  Using these ideas, they show in \cite{benamram02sums}
that the lower bound holds even if cells have infinite precision, but
the set of operations is restricted.

Miltersen et al.~\cite{miltersen94incremental} observe that there is a
trivial reduction from the partial-sums problem in $\mathbb{Z} /
2\mathbb{Z}$ to dynamic connectivity, implying an $\Omega(\lg n /
\lg\lg n)$ lower bound for the latter problem.  Independently, Fredman
and Henzinger \cite{fredman98connect} observe the same reduction, as
well as some more complex reductions applying to connectivity in plane
graphs and dynamic planarity testing.  Husfeldt and Rauhe
\cite{husfeldt03sums} show slightly stronger results using the
chronogram technique.  They prove that the lower bound holds even for
nondeterministic algorithms, and even in a promise version of the
problem in which the algorithm is told the requested sum to a $\pm 1$
precision.  These improved results make it possible to prove
reductions to various other problems \cite{husfeldt03sums,
husfeldt96sums}.

Alstrup, Husfeldt, and Rauhe \cite{alstrup98marked} give the only
previous improvement to the bounds of Fredman and Saks, by proving a
stronger trade-off of $t_q \lg t_u = \Omega(\lg n)$.  This bound is
the best trade-off provable by the chronogram technique.  However, it
still cannot improve beyond $\max\{t_u, t_q\} = \Omega(\lg n / \lg\lg
n)$.  The problem they considered was the partial-sums problem
generalized to trees, where a query asks for the sum of a root-to-leaf
path.  (This is a variation of the more commonly known marked-ancestor
problem.)  Their bound is tight for balanced trees; for arbitrary
trees, our lower bound shows that $\Theta(\lg n)$ is the best
possible.

Miltersen \cite{miltersen99survey} surveys the field of cell-probe
complexity, and advocates ``dynamic language membership'' problems as
a standardized framework for comparing lower bounds.  Given a language
$L$ that is polynomial-time decidable, the \emph{dynamic language
membership problem} for $L$ is defined as follows.  For any given $n$
(the problem size), maintain a string $w \in \{0,1\}^n$ under two
operations: flip the $i\th$ bit of $w$, and report whether $w \in L$.
Through its minimalism, this framework avoids several pitfalls in
comparing lower bounds.  For instance, it is possible to prove very
high lower bounds in terms of the number of cells in the problem
representation (which, misleadingly, is often denoted~$n$), if the
cells are large \cite{miltersen99survey}.  However, these lower bounds
are not very interesting because they assume exponential-size cells.
In terms of the number of bits in the problem representation, all
known lower bounds do not exceed $\Omega(\lg n / \lg\lg n)$.

Miltersen proposes several challenges for future research, two of
which we solve in this paper.  One such challenge was to prove an
$\Omega(\lg n)$ lower bound for the partial-sums problem.  Another
such challenge, listed as one of three ``big challenges'', was to
prove a lower bound of $\omega(\lg n / \lg\lg n)$ for a dynamic
language membership problem.  We solve this problem because dynamic
connectivity can be phrased as a dynamic language membership problem
\cite{miltersen99survey}.

\subsection{The Partial-Sums Problem in Other Models}

The partial-sums problem has been studied since the dawn of data
structures, and has served as the prototypical problem for the study
of lower bounds.  Initial efforts concentrated on algebraic models of
computation.  In the semigroup or group models, the elements of the
array come from a black-box (semi)group.  The algorithm can only
manipulate the $\Delta$ inputs through additions and, in the group
model, subtractions; all other computations in terms of the indices
touched by the operations are free.

In the semigroup model, Fredman \cite{fredman81sums} gives a tight
logarithmic bound.  However, this bound is generally considered weak,
because updates have the form $A[i] \gets \Delta$.  Because additive
inverses do not exist, such an update invalidates all memory cells
storing sums containing the old value of $A[i]$.  When updates have
the form $A[i] \gets A[i] + \Delta$, Yao \cite{yao85sums} proved a
lower bound of $\Omega(\lg n / \lg\lg n)$.  Finally, Hampapuram and
Fredman \cite{hampapuram98sums} proved an $\Omega(\lg n)$ lower bound
for this version of the problem; their bound holds even for the
offline problem.  In higher dimensions, Chazelle
\cite{chazelle97range} gives a lower bound of $\Omega((\lg n / \lg\lg
n)^d)$, which also holds even for the offline problem.

In the group model, the best previous lower bound of $\Omega(\lg n /
\lg\lg n)$ is by Fredman and Saks \cite{fredman89cellprobe}.  A tight
logarithmic bound (including the lead constant) was given by
\cite{fredman82sums} for the restricted class of ``oblivious''
algorithms, whose behavior can be described by matrix multiplication.
For the offline problem, Chazelle \cite{chazelle97range} gives a lower
bound of $\Omega(\lg\lg n)$ per operation; this is exponentially
weaker than the best known upper bound.  No better lower bounds are
known in higher dimensions.


\subsection{Upper Bounds for the Partial-Sums Problem}

An easy $O(\lg n)$ upper bound for partial sums is to maintain a
balanced binary tree with the elements of $A$ in the leaves, augmented
to store partial sums for each subtree.  A simple variation of this
scheme yields an implicit data structure occupying exactly $n$ memory
locations \cite{fenwick94sums}.  For the \func{sum} query, it is easy
to obtain good trade-offs.  Using trees with branching factor~$B$, one
can obtain $t_q = O(\log_B n)$ and $t_u = O(B \log_B n)$, or $t_q =
O(B\log_B n)$ and $t_u = O(\log_B n)$.  These bounds can be rewritten
as $t_q \lg \frac{t_u}{t_q} = O(\lg n)$, or $t_u \lg \frac{t_q}{t_u} =
O(\lg n)$, respectively, which matches our lower bound for the case
$\delta = \Theta(b)$, and for the group model.  For \func{select}
queries, one cannot expect to achieve the same trade-offs, because
even for a polynomial update time, there is a superconstant lower
bound on the predecessor problem \cite{beame02predecessor}.  Exactly
what trade-offs are possible remains an open problem.

Dietz \cite{dietz89sums} considers the partial-sums problem with
\func{sum} queries on a RAM, when $\delta = o(b)$.  He achieves $O(\lg
n / \lg\lg n)$ running times provided that $\delta = O(\lg\lg n)$.
Raman, Raman, and Rao \cite{raman01succinct} show how to support
\func{select} in $O(\lg n / \lg\lg n)$, again if $\delta = O(\lg\lg
n)$.  For $t_u = \Omega(\lg n / \lg\lg n)$, the same $\delta$, and
\func{sum} queries, they give a trade-off of $t_q = O(\log_{t_u} n)$.
They achieve the same trade-off for \func{select} queries, when
$\delta = 1$.  Hon, Sadakane, and Sung \cite{hon03sums} generalize the
trade-off for \func{select} when $\delta = O(\lg\lg n)$.  All of these
results do not scale well with $b$ or $\delta$ because of their use of
precomputed tables.

\subsection{Upper Bounds for Dynamic Connectivity}

For forests, Sleator and Tarjan's classic data structure for dynamic
trees \cite{sleator83dynamic} achieves an $O(\lg n)$ upper bound for
dynamic connectivity.  A simpler solution is given by Euler tour trees
\cite{henzinger99connect}.  This data structure can achieve a running
time of $t_q = O(\frac{\lg n}{\lg (t_u / t_q)})$, matching our lower
bound.

For general graphs, the first to achieve polylogarithmic time per
operation were Henzinger and King \cite{henzinger99connect}.  They
achieve $O(\lg^3 n)$ per update, and $O(\lg n / \lg\lg n)$ per query,
using randomization and amortization.  Henzinger and Thorup
\cite{henzinger97sampling} improve the update bound to $O(\lg^2 n)$.
Holm, de~Lichtenberg, and Thorup \cite{holm01connect} give a simple
deterministic solution with the same amortized running time: $O(\lg^2
n)$ per update and $O(\lg n / \lg\lg n)$ per query.  The best known
result in terms of updates is by Thorup \cite{thorup00connect},
achieving nearly logarithmic running times: $O(\lg n (\lg\lg n)^3)$
per update and $O(\lg n / \lg\lg\lg n)$ per query.  This solution is
only a factor of $(\lg\lg n)^3$ away from our lower bound.
Interestingly, all of these solutions are on our trade-off curve.  In
fact, for any $t_u = \Omega(\lg n (\lg\lg n)^3)$, Thorup's solution
can achieve $t_q = O(\frac{\lg n}{\lg (t_u / q_t)})$, showing that our
trade-off curve is optimal for this range of~$t_u$.

For plane graphs, Eppstein et al.~\cite{eppstein92planar} give a
logarithmic upper bound.  Plane graphs are planar graphs with a given
topological planar embedding, specified by the order of the edges
around each vertex.  Our lower bound holds for such graphs, proving
the optimality of this data structure.

\section{Models} \label{models}

The cell-probe model is a nonuniform model of computation.  The memory
is represented by a collection of cells. Operations are handled by an
algorithm which can read and write cells from the memory; all
computation is free, and the internal state is unbounded.  However,
the state is lost at the end of an operation. Because state is not
bounded, it can be assumed that all writes happen at the end of the
operation.  If cells have $b$ bits, we restrict the number of cells to
$2^b$, ensuring that a pointer can be represented in one cell.  This
restriction is a version of the standard \emph{transdichotomous
assumption} frequently made in the context of the word RAM, and is
therefore natural in the cell-probe model as well.

We extend the model to allow for nondeterministic computation, in the
spirit of \cite{husfeldt03sums}.  Boolean queries can spawn any number
of independent execution threads; the overall result is an accept
(``yes'' answer) precisely if at least one thread accepts.  The
running time of the operation is the running time of the longest
thread.  Rejecting threads may not write any cells; accepting threads
may, as long as all accepting threads write exactly the same values.
Because of this restriction, and because updates are deterministic,
the state of the data structure is always well-defined.

All lower bounds in this paper hold under Las Vegas randomization,
i.e., zero-error randomization.  We consider a model of randomization
that is particularly easy to reason about in the case of data
structures.  When the data structure is created, a fixed subset of,
say, $2^{b-1}$ cells is initialized to uniformly random values; from
that point on, everything is deterministic.  This model can easily
simulate other models of randomization, as long as the total running
time is at most $2^{b-1}$ (which is always the case in our lower
bounds); the idea is that the data structure maintains a pointer to
the next random cell, and increments the pointer upon use.  For
nondeterministic computation, all accepting threads increment the
pointer by the largest number of coins that could be used by a thread
(bounded by the running time).  Using this model, one can immediately
apply the easy direction of Yao's minimax principle
\cite{yao77minimax}.  Thus, for any given distribution of the inputs,
there is a setting of the random coins such that the amortized running
time, in expectation over the inputs, is the same as the original
algorithm, in expectation over the random coins.  Using the
nonuniformity in the model, the fixed setting of the coins can be
hardwired into the algorithm.

We also consider Monte Carlo randomization, i.e., randomization with
two-sided error.  Random coins are obtained in the same way, but now
the data structure is allowed to make mistakes.  We do not allow the
data structure to be nondeterministic.  In this paper, we are
concerned only with error probabilities of $n^{-\Omega(1)}$; that is,
the data structure should be correct with high probability.  Note that
by holding a constant number of copies of the data structure and using
independent coins, the exponent of $n$ can be increased to any desired
constant.  In the data-structures world, it is natural to require that
data structures be correct with high probability, as opposed to the
bounded-error restriction that is usually considered in complexity
theory.  This is because we want to guarantee correctness over a large
sequence of operations. In addition, boosting the error from constant
to $n^{-c}$ requires $O(\lg n)$ repetitions, which is usually not
significant for an algorithm, but is a significant factor in the
running time of a data-structure operation.

\section{Lower Bounds, Take One} \label{simple-sums}

In this section, we give the intuition behind our approach and detail
a simple form of it that allows us to prove an $\Omega(\lg n)$ lower
bound on the partial-sums problem when $\delta = \Theta(b)$, which is
tight in this case.  This proof serves as a warmup for the more
complicated results in Sections \ref{verify-sums} and \ref{highb}.

\subsection{General Framework}

We begin with the framework for our lower bounds in general terms.
Consider a sequence of data-structure operations $A_1, A_2,
\dots, A_m$, where each $A_i$ incorporates all information
characterizing operation $i$, i.e., the operation type and any
parameters for that type of operation. Upon receiving request $A_i$,
the data structure must produce an appropriate response. The
information gathered by the algorithm during a query (by probing certain
cells) must uniquely identify the correct answer to the query, and
thus must encode sufficient information to do so.

To establish the lower bounds of this paper, we establish lower bounds
for a simpler type of problem. Consider two adjacent intervals of
operations: $A_i, \dots, A_{j-1}$ and $A_{j}, \dots, A_k$.
At all times, conceptually associate with each memory cell a
\emph{chronogram}~\cite{fredman89cellprobe}, i.e., the index $t$ of
the operation $A_t$ during which the memory cell was last modified.
Now consider all read instructions executed by the data structure
during operations $A_j, \dots, A_k$ that access cells with a
chronogram in the interval $[i, j-1]$. In other words, we consider the
set of cells written during the time interval $[i, j-1]$ and read
during the interval $[j, k]$ before they are overwritten.
All \emph{information transfer} from time interval $[i, j-1]$ to
time interval $[j, k]$ must be encoded in such cells,
and must be executed by such cell writes and reads.
If the queries from the interval $[j,k]$ depend on updates from the
interval $[i,j-1]$, all the information characterizing this dependency
must come from these cell probes, because an update happening during
$[i,j-1]$ cannot be reflected in a cell written before time $i$.
The main technical part of our proofs is to establish a lower bound on the
amount of information that must be transferred between two time intervals,
which implies a corresponding lower bound on the number of cells that must be
written and read to execute such transfer.
Such bounds will stem from an encoding argument, in conjunction with a
simple information-theoretic analysis.

Next we show how to use such a lower bound on the information transfer
between two adjacent intervals of operations to prove
a lower bound on the data structural problems we consider.
Consider a binary tree whose leaves represent the entire sequence of
operations in time order.
Each node in the tree has an associated time interval of operations,
corresponding to the subtree rooted at that node.
We can obtain two adjacent intervals of operations by, for example,
considering the two nodes with a common parent.
For every node in the tree, we define the \emph{information transfer
through that node} to be the number of read instructions executed in the
subtree of the node's right child that read data written by (i.e.,
cells last written by) operations in the subtree of the node's left child.
The lower bound described above provides a lower bound on this information
transfer, for every node.
We combine these bounds into a lower bound on the number of cell probes
performed during the entire execution by simply summing over all nodes.

To show that this sum of individual lower bounds is indeed an overall
lower bound, we make two important points.  First, we claim that we
are not double counting any read instructions.  Any read instruction is
characterized by the time when it occurs and the time when the
location was last written.  Such a read instruction is counted by only
one node, namely, the lowest common ancestor of the read and write
times, because the write must happen in the left subtree of the node,
and the read must happen in the right subtree.  The second point
concerns the correctness of summing up individual lower bounds. This
approach works for the arguments in this paper, because all lower
bounds hold in the average case under the same probability
distribution for the operations.  Therefore, we can use linearity of
expectation to break up the total number of read instructions
performed on average into these distinct components.  Needless to say,
worst-case lower bounds could not be summed in this way.

The fact that our lower bounds hold in the average case of an input
distribution has another advantage: the same lower bound holds in the
presence of Las Vegas randomization. The proofs naturally allow the
running time to be a random variable, depending on the input. By the
easy direction of the minimax principle, a Las Vegas randomized data
structure can be converted into a deterministic data structure that on
a given random distribution of the inputs achieves the same expected
running time.

This line of argument has an important generalization that we use
for proving trade-off lower bounds.
Instead of considering a binary tree, we can consider a tree of
arbitrary degree.  Then we may consider the information transfer
either between any node and all its left siblings, or between any node
and all its right siblings. Neither of these strategies double counts
read instructions, because a read instruction is counted only for a node
immediately below the lowest common ancestor of the read and write times.

\subsection{An Initial Bound for the Partial-Sums Problem}

We are now ready to describe an initial lower bound for the
partial-sums problem, which gives a clear and concise materialization
of the general approach from the previous section. We will prove a
lower bound of $\Omega(\frac{\delta}{b}\lg n)$, which is tight
(logarithmic) for the special case of $\delta = \Theta(b)$. The bound
from this section only considers \func{sum} queries, and does not
allow nondeterminism.

It will be useful to analyze the partial-sums problem over an
arbitrary group with at least $2^{\delta}$ elements. Our proof will
not use any knowledge about the group, except the quantity~$\delta$.
Naturally, the data structure is allowed to know the group; in fact,
the data structure need only work for one arbitrary choice of group.
In particular, the lower bound will hold for the group
$\mathbb{Z} / 2^{\delta} \mathbb{Z}$, the group of
$\delta$-bit integers with addition modulo $2^{\delta}$. A solution
to the original partial-sums problem also gives a solution to the
problem over this group, as long as we can avoid overflowing a cell in
the original problem. To guarantee this, it suffices that $\delta +
\lg n < b$. By definition of the model, we always have $\lg n \leq b$
and $\delta \leq b$, so we can avoid overflow by changing only
constant factors.

We consider a sequence of $m = \Omega(\sqrt[3]{n})$ operations, where
$m$ is a power of two. Operations alternate between updates and
queries. We choose the index in the array touched by the operation
uniformly at random. If the operation is an update, we also choose the
value $\Delta$ uniformly at random. This notion of random updates and
queries remains unchanged in our subsequent lower bounds, but the
pattern of alternating updates and queries changes. Our lemmas do not
assume anything about which operations are updates or queries, making
it possible to reuse them later.

Our lower bound is based on the following lemma analyzing intervals of
operations.

\begin{lemma} \label{lem:sums}
Consider two adjacent intervals of operations such that the left interval
contains $L$ updates, the right interval contains $L$ queries, and overall
the intervals contain $O(\sqrt[3]{n})$ operations. Let $c$ be the
number of read instructions executed during the second interval that
read cells last written during the first interval. Then $E[c] =
\Omega(\frac{\delta}{b} L)$.
\end{lemma}

Before we embark on a proof of the lemma, we show how it implies our
logarithmic lower bound. As in the framework discussion, we consider a
complete binary tree with one leaf per operation. For every node $v$,
we analyze the information transfer through $v$, i.e., the read
instructions executed in the subtree of $v$'s right child that access
cells with a chronogram in the subtree of $v$'s left child. If $v$ is
on the $\frac{1}{3} \lg n$ bottommost levels, the conditions of the
lemma are satisfied, with $L$ being a quarter of the number of leaves
under $v$. Then, the information transfer through $v$ is $\Omega(L
\frac{\delta}{b})$ on average. As explained in the framework
discussion, we can simply sum these bounds for all nodes to get a
lower bound for the execution time. The information transfer through
all nodes on a single level is $\Omega(m \frac{\delta}{b})$ in expectation
(because these subtrees are disjoint). Over $\frac{1}{3} \lg n$ levels, the
lower bound is $\Omega(m \frac{\delta}{b} \lg n)$, or amortized
$\Omega(\frac{\delta}{b} \lg n)$ per operation.

\subsection{Interleaving Between Two Intervals}

The lower bound for two adjacent intervals of operations depends on
the interleaving between the indices updated and queried in the two
intervals. More precisely, we care about the indices $a_1, a_2, \dots$
touched by updates during the left interval of time, and the indices
$b_1, b_2, \dots$ queried during the right interval. By relabeling,
assume that $a_1 \le a_2 \le \cdots$ and $b_1 \le b_2 \le \cdots$.  We
define the \emph{interleaving number} $l$ to be the number of indices
$i$ such that, for some index $j$, $a_i < b_j \leq a_{i+1}$. In words,
the interleaving number counts transitions from runs of $a$'s to runs
of $b$'s when merging the two sorted lists of indices.

\begin{lemma}  \label{lem:interleave}
Consider two adjacent intervals of operations such that the left interval
contains $L$ updates, the right interval contains $L$ queries, and overall the
intervals contain $O(\sqrt[3]{n})$ operations. Then the interleaving
between the two intervals satisfies $E[l] = \Theta(L)$ and, with
probability $1 - o(1)$, no index is touched by more than one operation.
\end{lemma}

\begin{proof}
By the birthday paradox, the expected number of indices touched more
than once is at most $O((\sqrt[3]{n})^2) \cdot \frac{1}{n} =
O(n^{-1/3})$. By Markov's inequality, all indices are unique with
probability $1 - O(n^{-1/3})$. Because $l \le L$, it suffices to prove
the lower bound. We show $E[l \mid \textrm{all indices are unique}] =
\Omega(L)$. Because the condition is met with $\Omega(1)$ probability,
$E[l] = \Omega(L)$. Fix the set $S$ of $2L$ relevant indices
arbitrarily. It remains to randomly designate $L$ of these to be
updates from the left interval, and the rest of $S$ to be queries from
the right interval. Then $l$ is the number of transitions from updates
to queries, as we read $S$ in order. The probability that a transition
happens on any fixed position is $\frac{1}{4}$, so by linearity of
expectation, $E[l \mid S] = \Omega(L)$. Because this bound holds for
any~$S$, we can remove the conditioning.
\end{proof}

The following information-theoretic lemma will be used throughout the
paper, by comparing the lower bound it gives with upper bounds given
by various encoding algorithms. For an introduction to information
theory, we refer the reader to \cite{cover-infthy}. Remember that we
are considering the partial-sums problem over an arbitrary group with
at least $2^{\delta}$ elements.

\begin{lemma} \label{lem:entropy}
Consider two adjacent intervals of operations such that the left interval
contains $L$ updates, the right interval contains $L$ queries, and overall the
intervals contain $O(\sqrt[3]{n})$ operations. Let $G$ be the random
variable giving the indices touched by every operation, and giving the
$\Delta$ values for all updates except those in the left interval.
Let $S$ be the random variable giving all partial sums queried in the
right interval. Then $H(S \mid G) = \Omega(L\delta)$.
\end{lemma}

\begin{proof}
Fix $G=g$ to an arbitrary value, such that no index is touched twice
in the two intervals. Let $l$ be the interleaving between the two
intervals ($l$ is a function of $g$). Let $U$ denote the set of
indices updated in the left interval. By the definition of the
interleaving number, there must exist $l$ queries in the right
interval to indices $q_1 < q_2 < \cdots < q_l$ such that $U \cap
[q_{t-1} + 1, q_t] \neq \emptyset$ for each $t \ge 1$, where $q_0$ is
taken to be $-\infty$. Now let us consider the partial sums queried by
these $l$ queries, which we denote $S_1, S_2, \dots, S_l$. The terms of
these sums are elements of the array $A[1 \twodots n]$ at the time the
query is made. Some elements were set by updates before the first
interval, or during the second interval, so they are constants for
$G=g$.  However, each $S_t$ contains a random term in $[q_{t-1} + 1,
q_t]$, which comes from an update from the first interval. This element was
not overwritten by a fixed update from the second interval because,
by assumption, no index was updated twice. Then each $S_t$ will be a
random variable uniformly distributed in the group: even if we
condition on arbitrary values for each but one of the random terms,
the sum remains uniformly random in the group because of the existence of
inverses. Furthermore, the random variables will be independent,
because $S_t$ contains at least one random term that was not present
in any $S_r$ with $r < t$ (namely, the term in $[q_{t-1}+1, q_t]$).
Then $H((S_1, \dots, S_l) \mid G=g) = l\delta$. The variable $S$
entails $S_1, \dots, S_l$, so $H(S \mid G=g) \ge l\delta$. By Lemma
\ref{lem:interleave}, $E[l] = \Omega(L)$. Furthermore, with
probability $1-o(1)$, a random $G$ leads to no index being updated
twice in the two intervals, so the above analysis applies. Then $H(S
\mid G) = \Omega(L\delta)$.
\end{proof}

\subsection{Proof of Lemma \protect\ref{lem:sums}}

We consider two adjacent intervals of time, the first spanning
operations $[i,j-1]$ and the second spanning operations $[j,k]$.  We
propose an encoding for the partial sums queried in $[j,k]$ given the
value of $G$, and compare its size to the $\Omega(L\delta)$ lower
bound of Lemma \ref{lem:entropy}. Our encoding is simply the list of
addresses and contents of the cells probed in the right interval that
were written in the left interval. Thus, we are proposing an encoding
of expected size $E[c] \cdot 2b$ bits, proving that $E[c] = \Omega(L
\frac{\delta}{b})$. It should be noted that $c$ is a random variable,
because the algorithm can make different cell probes for different
update parameters.

To recover the partial sums from this encoding, we begin by running
the algorithm for the time period $[1,i-1]$; this is possible because
all operations before time $i$ are known given $G$. We then skip the
time period $[i,j-1]$ and run the algorithm for the time period
$[j,k]$, which will return the partial sums queried during this
time. To see why this is possible, notice that a read instruction
issued during time period $[j,k]$ falls into one of three categories,
depending on the time $t_w$ when the cell was written:

\begin{description}
\item[$t_w \geq j$:] We can recognize this case by maintaining a
  list of memory locations written during the simulation; the data is
  immediately available.
\item[$i \leq t_w < j$:] The contents of the memory location is
  available as part out encoding; we can recognize this case by
  examining the set of addresses in the encoding.
\item[$t_w < i$:] This is the default case, if we failed to satisfy
  the previous conditions. The contents of the cell is determined from
  the state of the memory upon finishing the first simulation up to
  time $i-1$.
\end{description}

\subsection{Obtaining Trade-Off Lower Bounds} \label{sec:trade-off}

We now show how our framework can be used to derive trade-off lower
bounds. In a nutshell, we consider instances where the cheaper
operation is performed more frequently, so that the total cost of
queries matches the total cost of updates. Then, we analyze the
sequence of operations by considering a tree with a higher branching
factor. 

Assume there exists a data structure with amortized expected running
times bounded by $t_u$ for updates and $t_q$ for queries. Our hard
instance consists of blocks of $t_u + t_q$ operations. Each block
contains $t_q$ updates and $t_u$ queries; the order inside a block is
irrelevant.
We generate the arguments to updates and queries randomly as before.
Let $B = 2\cdot \max\left\{ \frac{t_u}{t_q}, \frac{t_q}{t_u}\right\}$. We prove below
that the expected amortized cost of a block is $\Omega\left(\max\{t_u, t_q\}
\frac{\delta}{b} \log_B n\right)$. On the other hand, the expected amortized
cost of a block is at most $2 t_u t_q$. This implies $\frac{t_u
t_q}{\max\{t_u, t_q\}} = \Omega\left(\frac{\delta}{b} \log_B n\right)$, so
$\min\{t_u, t_q\} \cdot \lg \frac{\max\{t_u, t_q\}}{\min\{t_u, t_q\}} =
\Omega\left(\frac{\delta}{b} \lg n\right)$. This is the desired trade-off,
which is tight when $\delta = \Theta(b)$.

To prove the lower bound on blocks, consider a balanced $B$-ary tree
in which the leaves correspond to blocks. We let the total number of
blocks be $m = \Theta(\sqrt[6]{n})$. If $\max\{t_u, t_q\} =
\Omega(\sqrt[6]{n})$, our lower bound states that $\min\{t_u, t_q\} =
\Omega(1)$, so there is nothing to prove. Thus, we can assume $t_u +
t_q = O(\sqrt[6]{n})$, which bounds the number of operations in a
block. Then, the total number of operations is $O(\sqrt[3]{n})$,
satisfying one of the conditions of Lemma \ref{lem:sums}.

For the case $t_u \geq t_q$, we are
interested in the information transfer between each node and its left
siblings. The subtree of the node defines the right interval of
operations, and the union of the subtrees of all left siblings defines
the left interval. Let $L$ be the number of blocks in the right
interval. We make a claim only regarding nodes that are in the right
half of their parent's children. In this case, the number of blocks in
the left interval is at least $\frac{B}{2} L$. Then, the number of
queries in the right interval is $L t_u$, while the number of updates
in the left interval is at least $L \frac{t_u}{t_q} t_q = L t_u$. We
can then apply Lemma \ref{lem:sums}; having more updates in the left
interval cannot decrease the bound, because moving the beginning of
the left interval earlier can only increase the number of cell probes
that are counted. Therefore, the expected number of cell probes
associated with this node is $\Omega(L t_u \frac{\delta}{b})$. Now we
sum the lower bounds for all nodes on a level, and obtain that the
number of cell probes associated with that level is $\Omega(m t_u
\frac{\delta}{b})$. Summing over all levels, we get an amortized lower
bound of $\Omega(t_u \frac{\delta}{b} \log_B n)$ per block, as desired.

For the case $t_u < t_q$, we apply a symmetric argument. We analyze
the information transfer between any node, giving the left interval,
and all its right siblings, giving the right interval. For nodes in
the first half of their parent's children, the left interval contains
$L t_q$ updates, while the right interval contains at least $L t_q$
queries. By Lemma~\ref{lem:sums}, the expected number of cell probes
associated with this node is $\Omega(L t_q \frac{\delta}{b})$. Thus,
the number of cell probes associated with a level is $\Omega(m t_q
\frac{\delta}{b})$, and the amortized bound per block is $\Omega(t_q
\frac{\delta}{b} \log_B n)$.

\subsection{Refinements} \label{bitrev}

First note that our lower bounds so far depend only on randomness in
the update parameters $\Delta$, and not on randomness in the update or
query indices. Indeed, the value of $G$ is irrelevant, except for the
interleaving number that it yields. It follows that the logarithmic
lower bound and the trade-off lower bound are also true for
sequences of operations in which we fix everything except the $\Delta$
parameters, as long as such sequences have a high sum of the
interleaving numbers of each node. Our application of Lemma
\ref{lem:interleave} can be seen as a probabilistic proof that such
bad sequences exist.

The prototypical deterministic sequence with high total interleaving
is the bit-reversal permutation. For any $n$ that is a power of two,
consider the permutation $\pi : \{0,\dots, n-1\} \to \{0,\dots,n-1\}$
that takes $i$ to the integer obtained by reversing $i$'s $\log_2 n$ bits.
The corresponding access sequence consists of $n$ pairs of $\func{update}$ and
$\func{sum}$, the $i\th$ pair touching index $\pi(i)$.
The bit-reversal permutation underlies the Fast Fourier Transform algorithm.
It also gives an access sequence that takes $\Omega(\lg n)$ amortized time
for any binary search tree \cite{wilber89splay}. Finally, it was used to prove
an $\Omega(\lg n)$ bound for the partial-sums problem in the semigroup
model \cite{hampapuram98sums}. To see why this permutation has high
total interleaving, consider the following recursive construction. The
permutation $\pi'$ of order $2n$ is obtained from a permutation $\pi$ of
order $n$ by the rules: $\pi'(i) = 2\cdot \pi(i)$, $\pi'(i+n) = 2\cdot
\pi(i) + 1$, for $i \in \{0, 1, \dots, n-1\}$. Each level of the
recursion adds an interleaving of $n$ between the left and right
halves, so the total interleaving is $\Theta(n \lg n)$.

The fact that our lower bound holds for fixed sequences of operations
implies the same lower bound in the group model. A solution in the
group model handles every \func{update} and \func{sum} by executing a
sequence of additions on cells containing abstract elements from the
group. The cells touched by these additions depend only on the indices
touched by queries and updates, because the data structure treats the
group as a black box, and cannot examine the $\Delta$'s. So if we know
a priori the sequence of indices touched by queries and updates, we can
implement the same solution in the cell-probe model for the group
$\mathbb{Z} / 2^b \mathbb{Z}$; because the $\Delta$'s are unrestricted
elements of the group, $\delta = b$. The group additions can be
hard-wired into our solution for the cell-probe model through
nonuniformity, and cell probes are needed only to execute the actual
additions.

\subsection{Duality of Lower and Upper Bounds}

Recall the classic upper bound for the partial-sums problem.
We maintain a tree storing the elements of the array in order in the leaves.
Each node stores the sum of all leaves in its subtree.
An update adds $\Delta$ to the subtree sums along the root-to-leaf path of
the touched element.
A query traverses the root-to-leaf path of the element
and reports the sum of all subtrees to the left of the path.

Our lower bound can be seen as a dual of this natural algorithm. To
see this, we describe what happens when we apply the lower bound
analysis to the algorithm. We argue informally. Consider two intervals
of $2^k$ operations. The information transfer between the intervals is
associated with a node of height $k$ in the lower-bound tree. On the
other hand, the indices of the operations will form a relatively
uniformly spaced set of $O(2^k)$ indices. Thus, the distance in index
space between a query from the right interval and the closest update
from the left interval will usually be around $n / 2^k$. The algorithm's
tree passes information between the update and the query through the
lowest common ancestor of the two indices. Because of
the separation between the indices, this will usually be a node at
level around $\lg n - k$. Thus, we can say that our lower bound is
roughly an upside-down view of the upper bound. The information passed
through the $k\th$ level from the bottom of one tree is roughly associated
with the $k\th$ level from the top of the other tree.

\section{Handling Queries with Low Output Entropy} \label{verify-sums}

The lower bound technique as presented so far depends crucially on the
query answers having high entropy: the information transfer through a
node is bounded from below by the entropy of all queries from the
right subtree of the node. However, in order to prove lower bounds for
dynamic language membership problems (such as dynamic connectivity),
we need to be able to handle queries with binary answers. To prove
lower bounds for a pair of adjacent intervals, it is tempting to
consider the communication complexity between a party holding the
updates from the left interval, and a party holding the queries from
the right interval. Many bounds for communication complexity hold even
for decision problems, so queries with binary output should not be a
problem. However, a solution for the data structure does not really
translate well into the communication-complexity setting. The query
algorithm probes many cells, only a few of which (a logarithmic
fraction) are in the left interval. If the party with the right
interval communicates all these addresses, just to get back the answer
``not written in the left interval'' for most of them, the
communication complexity blows up considerably. One could also imagine
a solution based on approximate dictionaries, where the party holding
the left interval sends a sketch of the cells that were written,
allowing the other party to eliminate most of the uninteresting cell
probes. However, classic lower bounds for approximate dictionaries
\cite{carter78bloom} show that it is impossible to send a sketch that
is small enough for our purposes.  The solution developed in this
section is not based on communication complexity, although it can be
rephrased in terms of nondeterministic communication complexity. While
this solution is not particularly hard, we find it to be quite subtle.

\subsection{Setup for the Lower Bound} \label{sec:setuplb2}

Our approach is to construct hard sequences of operations that will
have a fixed response, and the data structure need only confirm that
the answer is correct. Such predictable answers do not trivialize the
problem: the data structure has no guarantee about the sequence of
operations, and the information it gathers during a query (by probing
certain cells) must provide a certificate that the predicted answer is
correct. In other words, the probed cells must uniquely identify the
answer to the query, and thus must encode sufficient information to do
so. As a consequence, our lower bounds hold even if the algorithm
makes nondeterministic cell probes, or if an all-powerful prover
reveals a minimal set of cells sufficient to show that a certain
answer to a query is correct.

The machinery developed in this section is also necessary in the case
of the partial-sums problem, if we want a lower bound for sequences of
\func{update} and \func{select} operations. Even though \func{select}
returns an index in the array, i.e., $\lg n$ bits, it is not clear how
more than one bit of information can be used for a lower-bound
argument. Instead, we consider a \func{verify-sum} operation, which is
given a sum $\Sigma$ and an index $i$, and tests whether the partial
sum up to $i$ is equal to $\Sigma$. In principle, this operation can be
implemented by two calls to \func{select}, namely by testing that $i =
\func{select}(\Sigma) = \func{select}(\Sigma - 1) + 1$.

Below we give a single lower-bound proof that applies to both the
partial-sums problem with verify, and dynamic connectivity. We
accomplish this by giving a proof for the partial-sums problem over
any group $G$ with at least $2^{\delta}$ elements, and then
specializing $G$ for the two problems we consider.

For the partial-sums problem with \func{select}, we use $G =
\mathbb{Z} / 2^{\delta} \mathbb{Z}$. This introduces a slight
complication, because \func{verify-sum} in modulo arithmetic can no
longer be implemented by a constant number of calls to
\func{select}. To work around this issue, remember that our lower
bound for \func{verify-sum} also holds for nondeterministic
computation. To implement $\func{verify-sum}(i, \Sigma)$
nondeterministically, we guess a $b$-bit quantity $\Sigma'$ such that
$\Sigma'\,\mathrm{mod}\,2^\delta = \Sigma$, and verify the old
condition $i = \func{select}(\Sigma') = \func{select}(\Sigma' - 1) +
1$.  We have implicitly assumed that \func{select} is deterministic,
which is natural because \func{select} does not return a binary
answer. Note that only one thread accepts, so there is no problem if
\func{select} updates memory cells (the updates made by the sole
accepting thread are the ones that matter).

For the dynamic-connectivity problem, we use $G = S_{\sqrt{n}}$,
i.e., the permutation group on $\sqrt{n}$ elements. Notice that now we
have $\delta = \sqrt{n} \lg \sqrt{n} - \Theta(\sqrt{n})$, a very large
quantity, unlike in the partial-sums problem where it was implied that
$\delta < b$. Our proof never actually assumes any particular relation
between $\delta$ and $b$.

\begin{figure*}
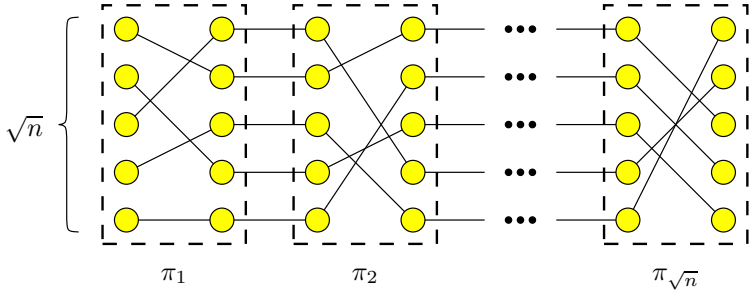

  \centering
  \ifpdf
    \input{multipermutation.pdf_t}
  \else
    \input{multipermutation.pstex_t}
  \fi
  \caption{Our graphs can be viewed as a sequence of permutation boxes
    (dashed).  The horizontal edges between boxes are in fact contracted
    in the actual graphs.}
  \label{fig:multiperm}
\end{figure*}

To understand the relation between this problem and dynamic
connectivity, refer to Figure \ref{fig:multiperm}.  We consider a
graph whose vertices form an integer grid of size $\sqrt{n}$ by
$\sqrt{n}$. Edges only connect vertices from adjacent columns. Each
vertex is incident to at most two edges, one edge connecting to a
vertex in the previous column and one edge connecting to a vertex in
the next column.  These edges do not exist only when they cannot
because the vertex is in the first or last column.  The edges between
two adjacent columns of vertices thus form a perfect matching in the
complete bipartite graph $K_{\sqrt n, \sqrt n}$, describing a
permutation of order $\sqrt{n}$. More precisely, point $(x,y_1)$ in
the grid is connected to point $(x+1, y_2)$ exactly when $\pi_x(y_1) =
y_2$ for a permutation~$\pi_x$.  Another way to look at the graph is
in terms of permutation networks. We can imagine that the graph is
formed by $\sqrt{n}$ horizontal wires, going between permutation
boxes. Inside each box, the order of all wires is changed arbitrarily.

Our graph is always the disjoint union of $\sqrt{n}$ paths.
This property immediately implies that the graph is
plane, because any embedding maintains planarity (though the edges
may have to be routed along paths with several bends).

The operations required by the partial-sums problem need to be
implemented in terms of many elementary operations, so they are are
actually ``macro-operations''. Macro-operations are of two types,
\func{update} and \func{verify-sum}, and all receive as parameters a
permutation and the index $x$ of a permutation box. To perform an
update, all the edges inside the named permutation box are first
deleted, and then reconstructed according to the new permutation. This
translates to $\sqrt{n}$ \func{delete}'s and $\sqrt{n}$ \func{insert}'s
in the dynamic connectivity world. Queries on box $x$ test that point
$(1,y)$ is connected to point $(x+1,\pi(y))$, for all $y \in \{1, 2,
\dots, \sqrt{n}\}$. This requires $\sqrt{n}$ connectivity queries. The
conjunction of these tests is equivalent to testing that the
composition of $\pi_1, \pi_2, \dots, \pi_x$ (the permutations
describing the boxes to the left) is identical to the given
permutation~$\pi$ --- the \func{verify-sum} in the partial-sums world.

As stated before, the lower bound we obtain is
$\Omega(\frac{\delta}{b} \lg n)$. For dynamic connectivity, we are
interested in $b = \Theta(\lg n)$, which is the natural word size for
this problem. As we saw already, $\delta = \Theta(\sqrt{n} \lg
n)$. Thus, our lower bound translates into $\Omega(\sqrt{n} \lg
n)$. This is a lower bound for the macro-operations, though, which are
implemented through $O(\sqrt{n})$ elementary operations. Therefore,
the lower bound for dynamic connectivity is $\Omega(\lg n)$, as
desired. The same calculation applies to the trade-off expressions,
which essentially means that the $\frac{\delta}{b}$ term should be
dropped to obtain the true bound for dynamic connectivity.

\subsection{Proof of the Lower Bound} \label{sec:verifylb}

As before, the sequence of operations alternates between \func{update}
or \func{verify-sum}. The index queried or updated is chosen uniformly
at random. If the operation is an \func{update}, we select a random
element of $G$ for the value $\Delta$. If the operation is
\func{verify-sum}, we give it the composition of the elements before
the queried index. This means that the data structure will be asked to
prove a tautology, involving the partial sum up to that index.

Because of this construction of the hard sequence, at least one
nondeterministic thread for each query should accept. For every random
input, let us fix one accepting thread for each operation. When we
mention cells that are ``read'', we mean cells read in this chosen
execution path; by definition of the model, writes are the same for
all accepting threads. As in the framework discussion, we are
interested in lower bounds for the information transfer between
adjacent intervals of operations. The following lemma is an analog of
Lemma \ref{lem:sums}.

\begin{lemma} \label{lem:verifysums}
Consider two adjacent intervals of operations such that the left interval
contains $L$ updates, the right intervals contains $L$ queries, and overall
the intervals contain $O(\sqrt[3]{n})$ operations. Let $w$ be the
number of write instructions executed during the first interval, and
let $r$ be the number of read instructions executed during the second
interval. Also let $c$ be the number of read instructions executed
during the second interval that read cells last written during the
first interval. Then $E[c] = \Omega\left(L \frac{\delta}{b}\right) -
O\left(\frac{E[s]}{b}\right)$, where $s = \lg \binom{r+w}{r}$.
\end{lemma}

Note that the lower bound of this lemma is weaker than that of Lemma
\ref{lem:sums}, because of the additional term $\frac{E[s]}{b}$.
Before we prove this lemma (in the next section), let us see that this
term ends up being inconsequential, and we can still prove the same
bounds and trade-offs.

Consider a sequence of $m = \Theta(\sqrt[3]{n})$ operations, and let
$T$ be the total running time of the data structure. We construct a
complete binary tree over these operations. Consider a node $v$ that
is a right child, and let $L$ be the number of leaves in its subtree.
By Lemma \ref{lem:verifysums}, we have $E[c] = \Omega\left(L
\frac{\delta}{b}\right) - O\left(\frac{E[s]}{b}\right)$,
where $c$ is the number of
cell probes associated with $v$. Note that $s = \lg \binom{r+w}{r}
\leq r + w$; thus, $s$ is bounded by the number of read instructions
in $v$'s subtree, plus the number of write instructions in the subtree
of $v$'s left sibling. Summing for all nodes on a level, $s$ counts
all read and write instructions at most once, so we obtain $E[\sum
c_i] = \Omega\left(m \cdot \frac{\delta}{b}\right) -
O\left(\frac{E[T]}{b}\right)$. We sum
up the lower bounds for each level to obtain a lower bound on
$E[T]$. We obtain that $E[T] = \Omega \left( m \cdot \frac{\delta}{b}
\lg n \right) - O\left( \lg n \cdot \frac{E[T]}{b} \right)$. Because
$\lg n \leq b$, this means that $E[T] = \Omega\left(m \cdot
\frac{\delta}{b} \lg n\right)$. This result implies an average-case
amortized lower bound per operation of $\Omega\left(\frac{\delta}{b} \lg
n\right)$.

To obtain trade-off lower bounds, we apply the same reasoning as in
Section \ref{sec:trade-off}. The only thing we have to do is verify
that the new term depending on $E[s]$ does not affect the end
result. When we sum the lower bounds for one level in the tree, we lose
a term of $O\left(\sum \frac{E[s_i]}{b}\right)$ compared to the old bound.
Here $i$ ranges over all nodes at that level. We must understand $\sum
s_i = \lg \prod \binom{r_i + w_i}{r_i}$ in terms of $T$, the total
running time for the entire sequence of operations.

The quantity $\prod \binom{r_i + w_i}{r_i}$ counts the total numbers
of ways to choose $r_i$ elements from a set $r_i + w_i$, where we have
a different set for each $i$. This is bounded from above by the number
of ways to choose $\sum r_i$ elements from a single set of $\sum (r_i
+ w_i)$ objects. Thus, $\sum s_i \le \lg \binom{\sum (r_i+w_i)} {\sum
r_i}$. Assume we have upper bounds $\sum r_i \le U_r$ and $\sum w_i \le
U_w$. Then, we can write $\binom{\sum (r_i + w_i)}{\sum r_i} \le
\binom{U_r + U_w}{\sum r_i} \le \binom{2(U_r + U_w)}{\sum r_i} \le
\binom{2(U_r + U_w)}{U_r}$. The last inequality holds because
$\binom{n}{k}$ increases with $k$ for $k \le \frac{n}{2}$. We entered
this regime by artificially doubling $U_r + U_w$. Because $r_i$ and
$w_i$ are symmetric, we also have $\binom{\sum (r_i + w_i)}{\sum r_i}
= \binom{\sum (r_i + w_i)}{\sum w_i} \le \binom{2(U_r + U_w)}{U_r}$.

Now we need to develop the upper bounds $U_r$ and $U_w$. For the case
$t_u \leq t_q$, our proof considered intervals formed by a node and
all its left siblings. Thus, $\sum r_i$ counts each read instruction
once, for the node it is under; so $\sum r_i \le T$. On the other
hand, $\sum w_i \le B\cdot T$, because a write instruction is counted
for every right sibling of its ancestor on the current level.  For
the case $t_u > t_q$, we consider intervals formed by a node, and all
its right siblings. Thus, $\sum w_i \le T$ and $\sum r_i \le B\cdot
T$.

Using these bounds, we see that $\sum s_i \le \lg \binom{2(B+1)T}{T} =
O(T \lg B)$. Because this upper bound holds in any random instance, it
also holds in expectation: $E[\sum s_i] = O(E[T] \lg B)$. So our lower
bound loses $O\left(\frac{E[T]\lg B}{b}\right)$ per level, which, over
all levels, sums to $O\left(\frac{E[T] \lg B}{b} \log_B n\right) =
O\left(E[T] \frac{\lg n}{b}\right)$. Because $\lg n \leq b$, our lower
bound on $E[T]$ is equal to the old lower bound minus $O(E[T])$.
Thus, we lose only a constant factor in the lower bound, and the
results of Section \ref{sec:trade-off} continue to hold.

\subsection{Proof of Lemma \protect\ref{lem:verifysums}}

The proof is an encoding argument, which is similar in spirit to the
proof of Lemma \ref{lem:sums}, but requires a few significant new
ideas. The difference from the previous proof is that the partial sums
that we want to encode are no longer returned by queries, but rather
they are given as parameters. Our strategy is to recover the partial
sums by simulating each query for all possible parameters, and see
which one leads to an accept. However, these simulations may read a
large number of cells, which we cannot afford in the encoding.
Instead, we add a new part to the encoding which enables us to stop
simulations that try to read cells we don't know. The difficulty is
making this new component of the encoding small enough.

As before, we consider two adjacent intervals of operations, the first
spanning $[i,j-1]$ and the second $[j,k]$. We propose an encoding for
the partial sums passed to the \func{verify-sum} operations during
$[j,k]$, given the variable $G$ defined in Lemma \ref{lem:entropy}. By
this lemma, such an encoding must have size $\Omega(L\delta)$ bits.

The encoder first simulates the entire execution of the data
structure. Each query is given the correct partial sum, so it must
accept. We choose arbitrarily one of the accepting threads. Consider
the following sets of cells, based on this computation history:

\begin{description}
\item[$W = $] cells which are updated by the data structure during the
  interval of time $[i,j-1]$, and never read during $[j,k]$.

\item[$R = $] cells which are read by the data structure during
 $[j,k]$ and their last update before the read happened before time
 $i$.

\item[$C = $] cells which are read by the data structure during
 $[j,k]$ and their last update before the read happened during
 $[i,j-1]$.
\end{description}

These are simple sets, so, for example, cells written multiple times
during $[i,j-1]$ are only included once in $W$. We have $|C| = c, |W|
\le w, |R| \le r$. Note that all of $c, |W|, w, |R|$ and $r$ are
random variables, because the data structure can behave differently
depending on the $\Delta$'s passed to the updates. We will give an
encoding for the queried partial sums that uses $O(b) + c \cdot 2b +
O(s)$ bits, where $s = \lg \binom{r+w}{r}$. Because the expected size
of our encoding must be $\Omega(L \delta)$, we obtain that $E[c] +
\frac{E[s]}{\Theta(b)} = \Omega\left(L \frac{\delta}{b}\right)$ and therefore
$E[c] = \Omega\left(L \frac{\delta}{b}\right) - O\left(\frac{E[s]}{b}\right)$.

Our encoding consists of two parts. The first encodes all information
about the interesting cell probes (the information transfer): for each
cell in $C$, we encode the address of the cell and its contents at
time $j$. This uses $O(b)$ bits to write the size of $C$, and $c \cdot
2b$ for the information about the cells. The second part is concerned
with the ``uninteresting'' cell probes, i.e.,~those in $R$. This
accounts for a covert information transfer: the fact that a cell was
\emph{not} written during $[i,j-1]$ is a type of information
transmitted to $[j,k]$. The part certifies that $W$ and $R$ are
disjoint, by encoding a set $S$, such that $R \subset S$ and $W
\subset \overline{S}$. We call $S$ a separator between $R$ and $W$. To
efficiently encode a separator, we need the following result:

\begin{lemma} \label{lem:separator}
For any integers $a,b,u$ with $a+b \leq u$, there exists a system of
sets $\mathbb{S}$ with $\lg |\mathbb{S}| = O(\lg\lg u + \lg
\binom{a+b}{a})$ such that, for all $A,B \subset \{1, 2, \dots, u\}$
with $|A| \le a, |B| \le b, A \cap B = \emptyset$, there exists an $S
\in \mathbb{S}$ satisfying $A \subset S$ and $B \subset \overline{S}$.
\end{lemma}

\begin{proof}
It suffices to prove the lemma for $|A| = a$ and $|B| = b$, because we
can simply add some elements from $\{1, 2, \dots, u\} \setminus (A
\cup B)$ to pad the sets to the right size. We use the probabilistic
method to show that a good set system exists. Select a set $S$
randomly, by letting every element $x \in \{1, 2, \dots, u\}$ be in
the set with probability $p = \frac{a}{a+b}$.  Then, for any pair
$A,B$, the probability that $A \subset S$ and $B \subset \overline{S}$
is $p^a (1-p)^b$. The system $\mathbb{S}$ will be formed of sets
chosen independently at random, so the probability that there is no
good $S$ for some $A$ and $B$ is $\left( 1 - p^a (1-p)^b
\right)^{|\mathbb{S}|} \leq \exp \left(-p^a (1-p)^b |\mathbb{S}|
\right)$. The number of choices for $A$ and $B$ is $\binom{u}{a}
\binom{u-a}{b} \leq u^{a+b}$. So the probability that there is no good
set in $\mathbb{S}$ for any $A,B$ is at most $u^{a+b} \exp \left(-p^a
(1-p)^b |\mathbb{S}| \right) = \exp \left( (a+b) \ln u - p^a (1-p)^b
|\mathbb{S}| \right)$. As long as this probability is less than $1$,
such a system $\mathbb{S}$ exists. So we want $(a+b) \ln u < p^a
(1-p)^b |\mathbb{S}| = \left(\frac{a}{a+b}\right)^a \left(
\frac{b}{a+b} \right)^b |\mathbb{S}|$. We want to choose a system of
size greater than $\frac{(a+b)^{a+b+1} \ln u}{a^a b^b}$. Then $\lg
|\mathbb{S}| = \Theta((a+b+1) \log_2 (a+b) + \lg\lg u - a\log_2 a -
b\log_2 b)$. Assume by symmetry that $a \leq b$. Then $\lg
|\mathbb{S}| = \Theta\left( \lg\lg u + a \lg \frac{b}{a} + a \cdot
\frac{b}{a}\log_2 \left( 1 + \frac{a}{b} \right) \right)$. Let $t =
\frac{b}{a}$; then $t \log_2 (1 + \frac{1}{t}) = \log_2 \left( (1 +
\frac{1}{t})^t \right) \to \log_2 e$ as $t \to \infty$. We then have
$\frac{b}{a} \log_2 \left( 1 + \frac{a}{b} \right) = \Theta(1)$, so
our result simplifies to $\lg |\mathbb{S}| = \Theta(\lg\lg u +
a\lg(b/a))$. It is well known that $a\lg(b/a) = \Theta(\lg
\binom{a+b}{a})$, for $a \leq b$.
\end{proof}

\medskip

We apply this lemma with parameters $r, w$ and $2^b$. First, we encode
$r$ and $w$, using $O(b)$ bits. The encoding and decoding algorithms
can simply iterate over all possible systems $\mathbb{S}$ for the
given $r$ and $w$, and choose the first good one (in the sense of the
lemma). Given this unique choice of a system, a separator between $R$
and $W$ is the index of an appropriate set in the system. This index
will occupy $O(\lg\lg (2^b) + \lg\binom{r+w}{r}) = O(\lg b + s)$ bits.

It remains to show that this information is enough to encode the
sequence of queried partial sums. We simulate the data structure for
the interval $[j,k]$, and prove by induction on time steps that all
cell writes made by these operations are correctly determined, and all
partial sums appearing in \func{verify-sum}'s are recovered. Updates
are easy to handle, because their parameters are known given $G$, and
they are deterministic. Thus, we can simply simulate the update
algorithm. We are guaranteed that all cells that are read and have a
chronogram in $[i,j-1]$ appear in $C$, so we can identify these cells
and recover their contents. All other cells have a known content given
$G$, so we can correctly simulate the update.

In the case of \func{verify-sum}, we do not actually know the sum
passed to it, so we cannot simply simulate the algorithm. Instead, we
try all partial sums that could be passed to the query, and for each
one try all possible execution paths that the data structure can
explore through nondeterminism. The cell probes made while simulating
such a thread fall in one of the following cases:

\begin{itemize}
\item the cell was written by the data structure after time $j$. This
  case can be identified by looking at the set of cells written during
  the simulation. By the induction hypothesis, we have correctly
  determined the cell's contents.

\item the cell is in $C$. We recover the contents from the encoding.

\item the cell is on $R$'s side of the separator between $R$ and
  $W$. Then, it was not written during $[i,j-1]$, and thus it has the
  old value before time $i$. Given $G$, everything is fixed before
  time $i$, so we know the cell's contents.

\item the cell is on $W$'s side of the separator. Then this thread of
  execution cannot be in the computation history chosen by the
  encoding algorithm. We abort the thread.
\end{itemize}

For each query, there exists a unique partial sum for which is should
accept. Furthermore, one accepting thread is included in the
computation history of the encoder. Thus, we identify at least one
thread which accepts and is not aborted because of the last case
above. Because the data structure is correct, all accepting threads
must be for the same partial sum, so we correctly identify the sum. By
definition of the nondeterministic model, the cell writes are
identical for all accepting threads, so we correctly determine the
cell writes, as well. 

It should be noted that, even though the size of the encoding only
depends on the characteristics of one accepting thread per query, the
separator allows us to handle an arbitrary number of rejecting
threads. All such threads (including all threads for incorrect partial
sums) are either simulated until they reject, or they are aborted.

\subsection{Handling Monte Carlo Randomization}

This section shows that the logarithmic lower bound for dynamic
connectivity is also true if we allow Monte Carlo randomization with
two-sided error probability at most $n^{-c}$, for constant $c$ (that
is, the data structure must be correct with high probability). The
idea is to make the decoding algorithm from the previous section use
only a polynomial number of calls to data structure operations. 

Assume for now that the data structure is deterministic. The previous
decoding algorithm simulates a large number of primitive connectivity
operations for every query. A partial sum is recovered by simulating
\func{verify-sum} for all possible sums. Remember that a partial sum
in the dynamic connectivity problem is a permutation in
$S_{\sqrt{n}}$, obtained by composition of the permutations up to a
certain column $k$. Thus, there are $(\sqrt{n})!$ partial sums to try
-- a huge quantity. However, we can recover the partial-sum
permutation by simulating at most $(\sqrt{n})^2$ \func{connected}
queries: it suffices to test connectivity of every point in the first
column with every point on the $k\th$ column. First, the decoder
simulates \func{connected} queries between the first node in column
one, and every node in column $k$. Exactly one of these queries was
executed by the encoder, so that query should accept. The other
queries will reject or be aborted. Now the writes made by the
accepting query are incorporated in the data structure. The decoder
continues to simulate query calls between the second node in column
one, and all nodes in column $k$, and so on.

Now assume that the data structure makes an error with probability at
most $n^{-c}$, for a sufficiently large constant $c$. We make the
encoding randomized; the random bits are those used to initialize the
memory of the data structure. We assume both the encoder and the
decoder receive the same random bits, so they can both simulate the
same behavior of the data structure. By the minimax principle, we can
fix that random bits if we are only interested in the expected size of
the encoding for a known input distribution (which is the case).

The decoding algorithm described above will work if all correct
queries accept, \emph{and} all incorrect queries would reject if they
were executed instead of the correct one. We can simulate the
execution of any query, or abort it only if it is not one of the
correct queries. So if all incorrect queries reject, their simulation
will either reject or be aborted. Because we only consider polynomial
sequences of operations, we simulate at most $\poly(n)$ queries
(including the incorrect ones). The probability that any of them will
fail is at most $n^{-c'}$, for some arbitrarily large constant $c'$
(depending of $c$). Because the decoder has the same coins as the
encoder, the encoder can predict whether the decoder will fail. Thus,
it can simply add one bit saying whether the old encoding is used
(when the decoder works), or the entire input is simply included in
the encoding (if the old decoder would fail). The expected size of the
encoding grows by at most $1 + n^{-c'} \cdot \poly(n) < 2$ for
sufficiently large $c'$. So the bounds of Lemma \ref{lem:verifysums}
remain the same. Then, the bounds and trade-offs derived for dynamic
connectivity hold even if the data structure answers correctly with
high probability.

\section{Handling a Higher Word Size} \label{highb}

For the partial-sums problem, it is natural and traditional to
consider the case $\delta = o(b)$. For ease of notation, we will let
$B = \frac{b}{\delta}$. For dynamic connectivity, our motivation comes
from external-memory models. For this problem, a ``memory cell'' is
actually an entire page, because that is the unit of memory that can be
accessed in constant time. In this case, $B$ is what is usually
referred to as ``page size''; the number of bits in a page is $b =
B\cdot \lg n$. For both problems, the lower bound we obtain is
$\Omega(\log_B n)$.

We note that the analysis from the previous sections gives a tight
bound on the number of bits that must be communicated: $\Omega(\delta
\lg n)$. Given that we can pack $b$ bits in a word, it is
straightforward to conclude that $\Omega(\frac{\delta}{b} \lg n) =
\Omega(\frac{\lg n}{B})$ read instructions must be performed. Our
strategy for achieving $\Omega(\frac{\lg n}{\lg B})$ is to argue that
an algorithm cannot make efficient use of all $b$ bits of a word, if
future queries are sufficiently unpredictable. Intuitively speaking,
if we need $\delta$ bits of information from a certain time epoch to
answer a query, and there are $t \cdot \frac{b}{\delta}$ possible
future queries that would also need $\delta$ bits of information from
the same epoch ($t > 1$), a cell probe cannot be very effective. No
matter what information the cell probe gathers, we have a probability
of at most $1/t$ that it has gathered all the information necessary
for a random future query, so with constant probability the future
query will need another cell probe. The reader will recognize the
similarity, at an intuitive level, with the round-elimination lemma
from communication complexity \cite{miltersen99asymmetric,sen-roundelim}.
Also note that our proof strategy hopelessly fails with
any deterministic sequence of indices, such as the bit-reversal
permutation. Thus, we are identifying another type of hardness hidden
in our problems.

Unfortunately, there are two issues that complicate our lower
bounds. The first is that, for dynamic connectivity, we need to go
beyond the \func{verify-sum} abstraction, and deal with
\func{connected} queries directly. To see why, remember that a
\func{verify-sum} macro-query accesses a lot of information
($\Theta(\sqrt{n} \lg n)$ bits) in a very predictable fashion,
depending on just one query parameter. Thus, we do not have the
unpredictability needed by our lower bound. The second complication is
that, for the partial-sums problem, we can handle
\func{verify-sum} only when $\delta = \Omega(b)$. When $\delta = o(\lg n)$,
the information per query is not enough to hide the cost of the
separators from Lemma~\ref{lem:separator}. However, we can still
obtain lower bounds for \func{sum} and \func{select}, without
nondeterminism, using a rather simple hack.

In Section \ref{sec:new-interv}, we describe a new analysis for
adjacent intervals of operations, which is the gist of our new lower
bounds. In Section \ref{sec:new-sums}, we show how this new lower
bound can be used for the partial-sums problems, whereas in Section
\ref{sec:new-connect}, we show how to apply it to dynamic
connectivity.

\subsection{A New Lower Bound for Adjacent Intervals} \label{sec:new-interv}

We now consider an abstract data-structure problem with two
operations, \func{update} and \func{query}. We do not specify what
\func{update} does, except that it receives some parameters, and
behaves deterministically based on those. A query receives two
parameters $i$ and $q$, and returns a boolean answer. We refer to $i$
as an index. For any admissible $i$, there exists a unique $q$ which
makes the query accept. The parameter $q$ is a $\delta$-bit value,
where $\delta$ is a parameter of the problem; we let $B =
b/\delta$. The implementation of \func{query} can be nondeterministic.
We assume that the hard instance of the problem comes from some random
distribution, but that the pattern of updates and queries is deterministic.
In the hard instance, each \func{query} receives the
$q$ which makes it accept. We will assume that the random features of each
operation are chosen by the distribution independently of the choices
for other operations. Though we do not really need this assumption,
it is true in both problems we consider, and assuming independence simplifies
exposition.

Now consider two intervals of operations $[i,j-1]$ and $[j,k]$, and
let $L$ be the number of queries in the second interval. We make an
information-theoretic assumption, which we will later prove for both
the partial-sums and dynamic-connectivity problems. To describe this
assumption, pick a random $t \in [j,k]$ such that the $t\th$ operation
is a query. Also pick a set $Q$ of $BL$ random queries that could have
been generated as query $t$. Now, imagine simulating each such query
starting with the state of the data structure at time $t-1$. Our
assumption is essentially that the correct $q$'s for all original
queries from $[j,k]$, plus the simulated queries in $Q$, have high
entropy. More specifically, let $Z$ be the random variable specifying
all updates outside $[i,j-1]$ and the indices for all queries,
including those in $Q$. We assume that the vector of $q$'s has entropy
$\Omega(BL\delta)$ given $Z$.

Let $w$ be the number of write instructions executed during the first
interval, and let $r$ be the number of read instructions executed during the
second interval. Also let $c$ be the number of read instructions
executed during the second interval that read cells last written
during the first interval. Under the assumptions above, we prove
$E[c] = \Omega(L) - O(\frac{s}{b})$, where $s = (B\cdot E[r]) \log_2
\left( 1 + \frac{E[w]}{B\cdot E[r]} \right)$.

We now outline the proof strategy. The probability that query $t$
reads at least one cell from $[i,j-1]$ is at most $\frac{E[c]}{L}$. If
$E[c]$ were small, so would this probability. That would mean that a
large fraction of the queries from $Q$ (random queries that could be
executed at time $t$) would not need to read any cell written during
$[i,j]$. On the other hand, the queries recover $\Omega(BL\delta) =
\Omega(Lb)$ bits of information about the updates in the left
interval. Because most queries don't need to read another cell, most of
this information must have already been recovered by the cell probes
made in $[j,t-1]$. There are at most $E[c]$ probes in expectation,
each reading $b$ bits, so the recovered information is not enough when
$E[c]$ is small.

\paragraph{Encoding Algorithm}
As the first step of the formal proof, we describe the algorithm
encoding the correct $q$'s. First simulate the entire execution of the
data structure, with the real query at time $t$. For each query,
include an arbitrary accepting thread in the computation
history. Based on this computation history, consider the following
sets:

\begin{description}
\item[$C = $] cells that are written during $[i,j-1]$ and read during
  $[j,k]$.
\item[$W = $] cells that are written during $[i,j-1]$ but not read
  during $[j,k]$.
\item[$R_1 = $] cells that are read during $[j,k]$, but never written
  during $[i,j-1]$.
\end{description}

Now simulate the queries in $Q$ starting from the state of the data
structure at time $t-1$. As before, we only pass correct parameters to
these queries. Call \emph{easy queries} the queries for which there
exists an accepting thread which does not read any cell in $W$; call
such a thread a \emph{good thread}. The rest of the queries are
\emph{hard queries}; let $h$ be the number of hard queries. Let $R_2$
be the union of the cells read by a arbitrary good thread of every
easy query, from which we exclude the cells in $C$. By definition of
easy queries, $R_2$ is disjoint from $W$. Let $R = R_1 \cup R_2$; $R$
is also disjoint from $W$.

The encoding has four parts:

\begin{enumerate}
\item encode $c$ and for each cell in $C$, the address and contents of
  the cell;

\item a separator (as given by Lemma \ref{lem:separator}) between $R$
  and $W$;

\item encode $h$, and the set of hard queries. The set takes $\lg
  \binom{|Q|}{h} = \lg \binom{BL}{h}$ bits.

\item the correct $q$ for each hard query, as an array of size
  $h$. This takes $h\delta$ bits.
\end{enumerate}

The third part of the encoding could be avoided for our current
problem, because the separator can be used to recognize hard
queries. However, we will later consider a variation in which we
discard the separator, and then encoding which queries are hard could
no longer be avoided.

\paragraph{Decoding Algorithm}
We now describe how to recover the correct $q$'s given $Z$ and the
previous encoding. By definition, a separator of $R$ and $W$ is also a
separator for $R_1$ and $W$. Given this separator and complete
information about $C$, we can simulate the real operations in the
second interval, as argued in Lemma \ref{lem:verifysums}, and recover
their correct $q$'s. Now we have to recover the correct parameters for
the queries in $Q$. For hard queries, this is included in the
encoding. For each easy query, each possible $q$, and all threads, we
try to simulate the thread starting with what we know about the data
structure at time $t-1$. Each cell that is probed falls in one the
following cases:

\begin{itemize}
\item the cell was written during $[j,t-1]$. Because we simulated the
  data structure in this interval, we can identify this condition and
  recover the cell contents.

\item the cell is in $C$. We recover the contents from the encoding.

\item the cell is on $R$'s side of the separator between $R$ and
  $W$. Then, it was not written during $[i,j-1]$, and we can recover
  the cell contents because $Z$ includes perfect information before
  time $i$.

\item the cell is on $W$'s side of the separator. Then, this thread of
  execution cannot be among the chosen good threads for the easy
  queries, so we abort it.
\end{itemize}

For the chosen good thread of an easy query, the encoder included its
probes outside of $C$ in the set $R_2$, so simulation of this thread
is never aborted. Thus, for each easy query, we find at least one
accepting thread, and recover the correct $q$.

\paragraph{Analysis}
We want to bound the size of the separator. We have $|W| \le w$,
$|R_1| \le r$, so it remains to bound $|R_2|$. In expectation over a
random $t$ and a random choice of the queries in the second interval,
the number of cells read by query $t$ is at most $\frac{E[r]}{L}$. We
are simulating a set of $BL$ queries as if they happened at time
$t$. In expectation, the total number of cell probes performed by
these is at most $BL \frac{E[r]}{L} = B \cdot E[r]$, which also bounds
$E[|R_2|]$. Then $E[|R|] \le E[|R_1|] + E[|R_2|] = O(B) E[r]$. To
specify the separator, we need $O(b)$ bits to write $|W|$ and $|R|$,
and then, by Lemma \ref{lem:separator}, $O\left(\lg b + \log_2 \binom{|W| +
|R|}{|R|}\right)$ bits for the index into the system of separators. The
total size is $O\left(b + |R| \log_2 \left(1 + \frac{|W|}{|R|}\right)\right)$
bits. The
function $(x,y) \mapsto x \log_2 (1 + \frac{y}{x})$ is concave, so the
expected size is upper bounded by moving expectations inside. Then,
the expected size of the separator is $O\left(b + (B \cdot E[r]) \lg \left(1 +
\frac{E[w]}{B\cdot E[r]}\right)\right) = O(b + s)$.

To analyze the rest of the encoding, we need to bound $h$. For a
random $t$, the expected number of cell probes from the first interval
that are made by query $t$ is at most $\frac{E[c]}{L}$. This means
that a random query at position $t$ is bad with probability at most
$\frac{E[c]}{L}$. Thus, $E[h] = BL \frac{E[c]}{L} = B\cdot
E[c]$. Explicitly encoding the correct $q$'s for the hard queries
takes $E[h] \delta = b\cdot E[c]$ bits in expectation. This is the
same as the space taken to encode the contents of cells in
$C$. Encoding which queries are hard takes space $O(b) + \lg
\binom{BL}{h} = O\left(b + h \lg \frac{BL}{h}\right)$.
The function $x \to x \lg
\frac{\gamma}{x}$ is concave for constant $\gamma$, so the expected
size is at most $O\left(b + E[h] \lg \frac{BL}{E[h]}\right) =
O\left(b + B\cdot E[c] \lg \frac{L}{E[c]}\right)$.

We have shown an upper bound of $O\left(E[c] b + s + B\cdot E[c] \lg
\frac{L}{E[c]}\right)$ on the expected total size of the encoding. Let
$\eps > 0$ be an absolute constant to be determined. If $E[c] \ge \eps
L$, there is nothing to prove. Otherwise, observe that $x \mapsto x
\log_2 (2 + \frac{\gamma}{x})$ grows with $x$ for constant $\gamma$,
so the last term of the encoding size becomes $O\left(B\eps L \lg
\frac{1}{\eps}\right)$.  The assumed lower bound on the size of the
encoding is $\Omega(BL\delta) = \Omega(bL)$, so we obtain $E[c] =
\Omega(L) - O\left(\frac{\eps}{\delta} L \lg \frac{1}{\eps}\right) -
O\left(\frac{s}{b}\right)$. Note that $\eps \lg \frac{1}{\eps}$ goes
to zero as $\eps$ goes to zero. Then, assuming $\delta \ge 2$, there
is an absolute constant $\eps$ such that the second term of the lower
bound is a constant fraction of the first term. We thus obtain $E[c] =
\Omega(L) - O\left(\frac{s}{b}\right)$.

\paragraph{Deterministic queries}
Now we consider a variation of our original problem, in which queries
are deterministic, and they return $q$, as opposed to verifying a
given $q$. The only change in our analysis is that we do not need the
separator. Indeed, each query can be simulated unambiguously, because
it only receives a known index, and it is deterministic. Then, the
separator term in our lower bound disappears, and we obtain $E[c] =
\Omega(L)$.

\subsection{The Partial-Sums Problem}  \label{sec:new-sums}

Our hard instance is the same as in Section~\ref{sec:trade-off}: we
consider blocks of $t_q$ random updates and $t_u$ queries to random
indices. We begin by showing a lower bound for two intervals based on
the analysis from the previous section. Let $c$, $r$, $w$ be as
defined in the previous section.

\begin{lemma} \label{lem:new-sumsint}
Consider two adjacent intervals of operations such that the left interval
contains $B\cdot L$ updates, the right interval contains $L$ queries, and
overall the intervals contain $O(\sqrt[3]{n})$ operations. The
following lower bounds hold:
\begin{itemize}
\item In the case of \func{sum} queries, $E[c] = \Omega(L)$.

\item In the case of \func{verify-sum} queries, $E[c] = \Omega(L) -
  O\left(\frac{s}{b}\right)$, where we define \mbox{$s = (B\cdot E[r]) \log_2
  \left( 1 + \frac{E[w]}{B\cdot E[r]} \right)$}.
\end{itemize}
\end{lemma}

\begin{proof}
This follows from the analysis in the previous section, as long as we
can show the information-theoretic assumption made there.
Specifically, we pick a query from the second interval, and imagine
simulating $BL$ random queries in its place. We need to show that the
partial sums of the original queries and these virtual queries have
entropy $\Omega(BL\delta)$, given the indices of all queries, and the
indices and $\Delta$ values for all queries outside the left interval
(the variable $Z$ from the previous section). To prove this, we apply
Lemma \ref{lem:entropy}. Because that lemma only deals with the partial
sums (a feature of the problem instance) and not with computation, it
doesn't matter that we are simulating the $BL$ queries at the same
time. The partial sums would be the same if the queries were ran
consecutively. Then, the lemma applies, and shows our entropy lower
bound. Note that the variable $G$ in Lemma \ref{lem:entropy} describes
all queries, including the simulated ones (which the lemma thinks are
consecutive). This is exactly the variable $Z$.
\end{proof}

We now show how to use this lemma to derive our lower bounds. Our
analysis is similar to that of Section \ref{sec:verifylb}, with two
small exceptions. The first is that there is an inherent asymmetry
between the left and right interval in Lemma \ref{lem:new-sumsint}.
Because of this, we can only handle the case $t_q = O(t_u)$. The second
change is that the definition of $s$ is somewhat different from that
in Lemma \ref{lem:verifysums}; roughly, $s$ is larger because $E[r]$
is multiplied by $B$. We will show lower bounds of the form $t_q \left(\lg
B + \lg \frac{t_u}{t_q}\right) = \Omega(\lg n)$.

We consider a balanced tree with branching factor $\beta = 2 B
\frac{t_u}{t_q}$, over $m = \Theta(\sqrt[6]{n})$ blocks. Because for
$\max\{t_q, t_u\} = \Omega(\sqrt[6]{n})$, our trade-off states
$\min\{t_q, t_u\} = \Omega(1)$, we may assume $t_u + t_q =
O(\sqrt[6]{n})$. Then there are $O(\sqrt[3]{n})$ operations in total,
as needed. We will consider right intervals formed by a node of the
tree, and left intervals formed by all its left siblings. The choice
of $\beta$ gives the right proportion of updates in the left interval
compared to queries in the right interval, for any node which is in
the right half of its siblings. Then, we can apply Lemma
\ref{lem:new-sumsint}.

First, consider the case of \func{sum} queries, so there is no term
depending on $s$. Note that the $\Omega(L)$ term is linear in the
number of queries, so summing it up over the entire tree yields a
lower bound on the total time of $E[T] = \Omega(m t_u \log_{\beta}
n)$. By the definition of the blocks, $E[T] = m \cdot 2t_u t_q$, so
$t_q = \Omega(\log_{\beta} n)$, which is equivalent to $t_q \left(\lg B +
\lg \frac{t_u}{t_q}\right) = \Omega(\lg n)$.

Now we consider nondetermistic \func{verify-sum} queries, assuming
$\delta = \Omega(\lg n)$. There is a new term in the lower bound on
$E[T]$, given by the sum of the $s$ terms over all nodes. First
consider the sum for all nodes on one level: $\sum s_i = \sum (B\cdot
E[r_i]) \log_2 \left(1 + \frac{E[w_i]} {B\cdot E[r_i]} \right)$. We
have $\sum r_i \le T$, because each read is counted for the node it is
under, and $\sum w_i \le \beta T$, because each write is counted for
the siblings of the node it is under. These inequalities must also
hold in expectation, so $\sum B\cdot E[r_i] \le B\cdot E[T]$ and $\sum
E[w_i] \le \beta E[T]$. Because the function $(x,y) \mapsto x \log_2 (1
+ \frac{y}{x})$ is concave, $\sum s_i$ is maximized when $E[r_i]$ and
$E[w_i]$ are equal. Then $\sum s_i \le B \cdot E[T] \log_2 \left( 1 +
\frac{\beta E[T]}{B \cdot E[T]} \right) = O(E[T] B \lg(t_u /
t_q))$. When we sum this over $O(\log_{\beta} n)$ levels, we obtain
$E[T] \cdot O\left(B \lg(t_u / t_q) \frac{\lg n}{\lg(B t_u / t_q)}\right) =
E[T] \cdot O(B \lg n)$.

Thus, our overall lower bound becomes $E[T] = \Omega(m t_u \log_{\beta}
n) - O(\frac{1}{b} E[T] B\lg n)$. Expanding $B$, $E[T] = \Omega(m t_u
\log_{\beta} n) - O(E[T] \frac{\lg n}{\delta})$. For $\delta =
\Omega(\lg n)$, we obtain $E[T] = \Omega(m t_u \log_{\beta} n)$, which
is the same lower bound as for \func{sum} queries. This bound holds
for \func{verify-sum} queries, even with nondeterminism, and, as shown
in Section~\ref{sec:setuplb2}, also for \func{select} queries.

For the case $\delta = O(\lg n)$, we can obtain the same lower bound
on \func{select} by a relatively simple trick. This means that the
trade-off lower bound for \func{select} holds for any $\delta$, though
we cannot prove it in general for \func{verify-sum}. The trick is to
observe that for small $\delta$ (e.g.~$\delta < \frac{1}{3} \lg n$),
we can stretch (polynomially) an instance of \func{sum} into an
instance of \func{select}. Because we already have a lower bound for
\func{sum}, a lower bound for \func{select} follows by reduction.

Consider the \func{sum} problem on an array $A[0 \twodots
\sqrt[3]{n}-1]$, where each element has $\delta$ bits. This implies $0
\le A[i] < \sqrt[3]{n}$, and any partial sum is less than
$n^{2/3}$. Now we embed $A$ into an array $A'[0 \twodots n-1]$ by
$A'[i \cdot n^{2/3}] = A[i]$. The $n^{2/3} - 1$ spacing positions
between elements from $A$ are set to $1$ in the initialization phase,
and never changed later. An update in $A$ translates into an update in
$A'$ in the appropriate position. Now assume we want to find $\sigma =
\sum_{i=0}^k A[i]$. We run $\func{select}((k+1) (n^{2/3} - 1))$ in
$A'$. We have $\sum_{j = 0}^{t + k \cdot n^{2/3}} A[j] = t + \sigma +
k (n^{2/3} - 1)$, for any $t < n^{2/3}$. Then, if $\func{select}$
returns $t + k \cdot n^{2/3}$, we know that $t+\sigma = n^{2/3}-1$, so
we find $\sigma$.

\subsection{Dynamic Connectivity} \label{sec:new-connect}

The graph used in the hard sequence is the same as the one before
(Figure \ref{fig:multiperm}): $\sqrt{n}$ permutation boxes, each
permuting $\sqrt{n}$ ``wires''. Let $t_u$ be the running time of an
update (edge insertion or deletion), and $t_q$ the running time of a
query. We only handle $t_q \le t_u$. Our hard sequence consists of
blocks of operations. Each block begins with a macro-update: for an
index $k$ (chosen as described below), remove all edges in the $k\th$
permutation box, and insert edges for a random permutation. Then, the
block contains $\frac{t_u}{t_q} \sqrt{n}$ \func{connected}
queries. Each query picks a random node in the first column and a
random index $k$, and calls \func{connected} on the node in the first
column and the node on the $k\th$ column which is on the same
path. This means that all queries should be answered in the
affirmative; the information is contained in the choice of the node
from the $k\th$ column.

We still have to specify the sequence of indices of the
macro-updates. We use a deterministic sequence to ensure that updates
which occur close in time touch distant indices. This significantly
simplifies the information-theoretic analysis. Our hard sequence
consists of exactly $\sqrt{n}$ block. Each macro-update touches a
different permutation box; the order of the boxes is given by the
bit-reversal permutation (see Section \ref{bitrev}) of order
$\sqrt{n}$. Now consider a set of indices $S = \{ i_1, i_2, \dots \}$
sorted by increasing $i_j$. We say $S$ is \emph{uniformly spaced} if
$i_{j+1} - i_j = \sqrt{n} / (|S| - 1)$ for every $j$.

\begin{lemma} \label{lem:new-connect}
Consider two adjacent intervals of operations, such that the second
one contains $L$ queries, and the indices updated in the first
interval contain a uniformly-spaced subset of cardinality $\Theta(BL /
\sqrt{n})$. Then $E[c] = \Omega(L) - O\left(\frac{s}{b}\right)$, where $s =
B\cdot E[r] \lg \left(1 + \frac{E[w]} {B\cdot E[r]} \right)$.
\end{lemma}

\begin{proof}
This lemma follows from Section \ref{sec:new-interv}, if we show the
information-theoretic assumption used there. For our problem, $\delta
= \Theta(\lg n)$.  Imagine picking a random query from the right
interval and simulating $BL$ random queries in its place. The
variable $Z$ denotes the random choices for all queries, and for
updates outside the left interval. We need to show that the entropy of
the correct parameters for all queries in the right interval,
including the simulated ones, given $Z$, is $\Omega(BL \lg n)$.

Remember that all updates are to different boxes, so an update is
never overwritten. For this reason, our proof will not care about the
precise order of updates and queries in the right interval, and there
will be no difference between the real and simulated queries. Let the
uniformly-spaced set of update indices be $S = \{ i_1, i_2, \dots \}$.
We let $B_j$ be the set of queries from the right interval (either
real or simulated) whose random column is in $[i_j, i_{j+1} - 1]$. For
notational convenience, we write $H(B_j)$ for the entropy of the
correct parameters to the set of queries $B_j$. Basic information
theory states that $H(\bigcup_j B_j \mid Z) = \sum_j H(B_j \mid B_1,
\dots, B_{j-1}, Z)$. Thus, to prove our lower bound, it suffices to
show $H(B_j \mid B_1, \dots, B_{j-1}, Z) = \Omega(\sqrt{n} \lg n)$ for
all $j$.

Let $Z_j$ be a random variable describing $Z$ and, in addition, the
random permutations for all updates in the left interval with indices
below $i_j$. Also, if there are any updates with indices in $[i_j + 1,
i_{j+1} - 1]$, include their permutation in $Z_j$ (these are updates
outside the uniformly-spaced set). Note that $H(B_j \mid B_1, \dots,
B_{j-1}, Z) \ge H(B_j \mid Z_j)$, because conditioning on $Z_j$ also
fixes the correct parameters for queries in $B_1, \dots, B_{j-1}$.

Now let us look at a query from $B_j$. The query picks a random node
in the first column. All permutations before column $i_j$ are fixed
through $Z_j$, so we can trace the path of the random node until it
enters box $i_j$. Assume we have the correct parameter of the query,
i.e., the node from column $k$ to which the initial node is
connected. Permutations between column $i_j$ and $i_{j+1}$ are also
fixed by $Z_j$, so we can trace back this node until the exit of box
$i_j$. Thus, knowing the correct parameter is equivalent to knowing
some point values of the permutation $i_j$. As long as the nodes
chosen in the first column are distinct, we will learn new point
values. If we query $d$ distinct point values of the random
permutation, the entropy of the correct parameters is $\Omega(d\lg
n)$, for any $d$.

Now imagine an experiment choosing the queries sequentially. This
describes a random walk for $d$. In each step, $d$ may remain constant
or it may be incremented. Becuase of the uniform spacing, the probability
that a query ends up in $B_j$ is $\Omega(\frac{\sqrt{n}} {BL})$. If $d
\le \sqrt{n} / 2$, with probability at least a half, the node chosen
in the first column is new. Then, for $d \le \sqrt{n} / 2$, the
probability that $d$ is incremented in $\Omega(\frac{\sqrt{n}}{BL})$.
We do $BL$ independent random steps, and we are interested in the
expected value of $d$ at the end. The waiting time until $d$ is
incremented is $O(\frac{BL}{\sqrt{n}})$. For a sufficiently small
constant $\eps$, the expected time until $d$ reaches $\eps \sqrt{n}$
is $\frac{1}{2} BL$. Then, with probability at least a half, $d \ge
\eps\sqrt{n}$ after $BL$ steps. This is implies the expected value of
$d$ after $BL$ steps is $\Omega(\sqrt{n})$, so $H(B_j \mid Z_j) =
\Omega(\sqrt{n} \lg n)$.
\end{proof}

To use this lemma, we construct a tree with a branching factor $\beta
\ge 2 B\frac{t_u}{t_q}$, rounded to the next power of two. The right
interval is formed by a node, and the left interval by the node's left
siblings. We only consider the case when the node is among the right
half of its siblings. Now we argue there is a uniformly-spaced subset
among the indices updated in the left interval. Note that these
include all indices from the first half of siblings. Because $\beta$ is
a power of two, a root-to-leaf path in the tree is tracing a bit
representation of the leaf's index, in chunks of $\log_2 \beta$
bits. Because update indices are the reverse of the leaf's index, all
the leaves in the subtrees of the first half of the children have the
same low order bits in the indices. On the other hand, the high order
bits assume all possible values. So the indices from the first half of
the children are always a uniformly-spaced subset of indices.

Now we can apply Lemma \ref{lem:new-connect}, and we sum over all
nodes of the tree to obtain our lower bound. By the analysis in the
previous section, the sum of the $s$ terms only changes the bound by a
constant factor. The $\Omega(L)$ term of the lower bound is linear in
the number of queries, so summing over all levels we obtain $E[T] =
\Omega(\sqrt{n} \cdot \frac{t_u}{t_q}\sqrt{n} \cdot \log_{\beta}
n)$. Because $E[T] = \sqrt{n} \left(t_u \sqrt{n} + t_q \frac{t_u}{t_q}
\sqrt{n}\right) = O(\sqrt{n}\cdot t_u \sqrt{n})$, we obtain $t_q =
\Omega(\log_{\beta} n)$, which is our desired lower bound.

\section{Upper Bounds for the Partial Sums Problem} \label{upperbnds}

As mentioned before, our partial-sums data structure can support a
harder variant of updates. We will allow the $A[i]$'s to be arbitrary
$b$-bit integers, while $\func{update}(i, \Delta)$ implements the
operation $A[i] \gets A[i] + \Delta$, where $\Delta$ is a $\delta$-bit
(signed) integer.

Our data structure is based on a balanced tree of branching factor $B$
(to be determined) with the elements of the array $A[1 \twodots n]$ in
the leaves. Assume we pick $B$ such that we can support constant-time
operations for the partial-sums problem in an array of size $B$. Then,
we can hold an array of size $B$ in every node, where each element is
the total of the leaves in one of the $B$ subtrees of our node. All
three operations in the large data structure translate into a sequence
of operations on the small data structures of the nodes along a
root-to-leaf path. Thus, the running time is $O(\log_B n)$. We will
show how to handle $B = \Theta(\min\{ b/\delta, b^{1/5} \})$. Then
$\lg B = \Theta(\lg (b/\delta))$, which implies our upper bound.

It remains to describe the basic building block, i.e., a constant-time
solution for arrays of $B$ elements. We now give a simple solution for
\func{update} and \func{sum}. In the next section, we develop the
ideas necessary to support \func{select}. We will conceptually
maintain an array of partial sums $S[1 \twodots B]$, where $S[k] =
\sum_{i=1}^k A[i]$. To make it possible to support \func{update} in
constant time, we maintain the array as two separate components, $V[1
\twodots B]$ and $T[1 \twodots B]$, such that $S[i] = V[i] +
T[i]$. The array $V$ will hold values of $S$ that were valid at some
point in the past, while more recent updates are reflected only in
$T$. We can use Dietz's incremental rebuilding scheme
\cite{dietz89sums} to maintain every element of $B$ relatively
up-to-date: on the $t\th$ \func{update}, we set $V[t \bmod B] \gets
V[t \bmod B] + T[t \bmod B]$ and $T[t \bmod B] \gets 0$.  This scheme
guarantees that every element in $T$ is affected by at most $B$
updates, and thus is bounded in absolute value by $B \cdot
2^{\delta}$.

The key idea is to pack $T$ in a machine word. We represent each
$T[i]$ by a range of $O(\delta + \lg n)$ bits from the word, with one
zero bit of padding between elements.  Elements in $T$ can also be
negative; in this case, each value will be represented in the standard
two's complement form on its corresponding range of bits. Packing $T$
in a word is possible as long as $B = O\left( \frac{b}{\delta + \lg b}\right)$.
We can read and write an element of $T$ using a constant number of
standard RAM operations (bitwise boolean operations and shift
operations).

To complete our solution, we need to implement \func{update} in
constant time. Using the packed representation, we can add a given
value to all elements $V[i]$, $i \geq k$, in constant time. Refer to
Figure \ref{bits}. First, we create a word with the value to be added
appearing in all positions corresponding to the elements of $V$ that
need to be changed. We can compute this word using a multiplication by
an appropriate binary pattern. The result is then added to the packed
representation of $V$; all the needed additions are performed in one
step, using word-level parallelism. Because we are representing
negative quantities in two's complement, additions may carry over, and
set the padding bits between elements; we therefore force these buffer
bits to zero using a bitwise \func{and} with an appropriate constant
mask.

\begin{figure*}[t]
\centering
\begin{tabular}{@{} c @{\hspace{1ex}} c @{\hspace{1ex}} c @{\hspace{1ex}}
    c @{\hspace{1ex}} c @{\hspace{1ex}} c @{\hspace{1ex}} c
    @{\hspace{1ex}} c @{\hspace{1ex}} c | l @{}}

$V[4]$ & 0 & $V[3]$ & 0 & $V[2]$ & 0 & $V[1]$ & 0 & $V[0]$ 
      & old packed representation of $V$ \\
\hline

00001 & 0 & 00001 & 0 & 00001 & 0 & 00001 & 0 & 00001 
      & constant pattern \\

00001 & 0 & 00001 & 0 & 00001 & 0 & 00000 & 0 & 00000
      & shift right, then left by same amount \\

 & & & & & & & & $\Delta$
      & argument given to \func{update} \\

$\Delta$ & 0 & $\Delta$ & 0 & $\Delta$ & 0 & 00000 & 0 & 00000
      & multiply the last two values \\

$V'[4]$ & ? & $V'[3]$ & ? & $V'[2]$ & ? & $V[1]$ & ? & $V[0]$
      & add to the packed representation of $V$ \\

11111 & 0 & 11111 & 0 & 11111 & 0 & 11111 & 0 & 11111
      & constant cleaning pattern \\
\hline

$V'[4]$ & 0 & $V'[3]$ & 0 & $V'[2]$ & 0 & $V'[1]$ & 0 & $V'[0]$
      & final value of $V$, obtained by bitwise \func{and} \\
\end{tabular}

\caption{Performing $\func{update}(2, \Delta)$ at the word level. Here
  $V$ has 5 elements, each 5 bits long.}
\label{bits}
\end{figure*}

\subsection{Selecting in Small Arrays}

To support \func{select}, we use the classic result of Fredman and
Willard \cite{fredman93fusion} that forms the basis of their
fusion-tree data structure. Their result has the following black-box
functionality: for $B = O(b^{1/5})$, we can construct a data structure
that can answer successor queries on a static array of $B$ integers in
constant time. As demonstrated in \cite{andersson99fusion}, the lookup
tables used by the original data structure can be eliminated, if we
perform a second query in the sketch representation of the array. The
data structure can then be constructed in $O(B^4)$ time.

As before, we break partial sums into the arrays $V$ and $T$. We store
a fusion structure that can answer successor queries in $V$. Because
the fusion structure is static, we abandon the incremental rebuilding
of $V$, in favor of periodic global rebuilding. By the standard
deamortization of global rebuilding \cite{marks-cgbook}, we can then
obtain worst-case bounds. Our strategy is to rebuild the data
structure completely every $B^4$ operations: we set $V[i] \gets V[i] +
T[i]$ and $T[i] \gets 0$, for all $i$, and rebuild the fusion
structure over~$V$.  While servicing a \func{select} that doesn't
occur immediately after a rebuild, the successor in $V$ found by the
fusion structure might not be the appropriate answer to the
\func{select} query, because of recent updates.  We will describe
shortly how the correct answer can be computed by also examining the
array $T$; the key realization is that the real successor must be
close to the successor in $V$ in terms of their partial sums.

Central to our solution is the way we rebuild the data structure every
$n^4$ operations. We begin by splitting $S$ into runs of elements
satisfying $S[i+1] - S[i] < B^4 \cdot 2^{\delta}$; recall that we must
have $S[i] < S[i+1]$ for the \func{select} problem. We denote by
$\rep(i)$ the first element of the run containing $i$ (the
representative of the run); also let $\len(i)$ be the length of the run
containing $i$.  Each of these arrays can be packed in a word, because
we already limited ourselves to $B = O(b^{1/5})$. Finally, we let
every $V[i] \gets V[\rep(i)]$ and $T[i] \gets S[i] - V[\rep(i)]$.
Conveniently, $T$ can still be packed in a word. Indeed, the value
stored in an element after a rebuild is at most $B \cdot \left( B^4
\cdot 2^{\delta} \right)$, and it can subsequently change by less than
$B^4 \cdot 2^{\delta}$. Therefore, it takes $O(\lg B + \delta)$ bits
to represent an element of $C$, so we only need to impose the
condition that $B = O( \min\{ b / \delta, b^{1/5} \})$.

It remains to show how $\func{select}(\sigma)$ can be answered. Let
$k$ denote the successor in $V$ identified by the fusion structure; we
have $V[k-1] < \sigma \leq V[k]$.  We know that $k$ is the
representative of a run, because all elements of a run have equal
values in~$V$. By construction, runs are separated by gaps of at least
$B^4 \cdot 2^{\delta}$, which cannot be closed by $B^4$ updates.
Thus, the answer to the query must be either an index in the run
starting at $k$, or an index in the run ending at $k-1$, or exactly
equal to $k+\len(k)$. We can distinguish between these cases in
constant time, using two calls to \func{sum} followed by
comparisons. If we identify the correct answer as exactly $k+\len(k)$,
we are done.

Otherwise, assume by symmetry that the answer is an index in the run
starting at $k$. Because elements of a run have equal values of $V$,
our task is to identify the unique index $i$ in the run satisfying
$T[i-1] < \sigma - V[k] \leq T[i]$. Now we can employ word-level
parallelism to compare all elements in $T$ with $\sigma - V[k]$ in
constant time. This is similar to a problem discussed by Fredman and
Willard \cite{fredman93fusion}, but we must also handle negative
quantities. The solution is to subtract $\sigma - V[k]$ in parallel
from all elements in $T$; if elements of $T$ are oversized by 1 bit,
we can avoid overflow. The sign bits of every element then give the
results of the comparisons. The answer to the query can be found by
summing up the sign bits corresponding to elements in our run, which
indicates how many elements in the run were smaller than $\sigma -
V[k]$. Because these bits are separated by more than $\lg b$ zeroes,
we can sum them up using a multiplication with a constant pattern, as
described in \cite{fredman93fusion}.

\section{Reductions To Other Dynamic Graph Problems} \label{reductions}

It is relatively easy to dress up dynamic connectivity as other
dynamic graph problems, obtaining logarithmic lower bounds for
these. Most problems on undirected graphs admit polylogarithmic
solutions, so such lower bounds are interesting. The problems discussed in
this section are only meant as examples, and not as an exhaustive list
of possible reductions.

\subsection{Connectivity of the Entire Graph}
The problem is to maintain a dynamic graph along with the answer to
the question ``is the entire graph connected?''. We obtain a lower
bound of $\Omega(\lg n)$ even for plane graphs, which implies the same
lower bound for counting connected components. The dynamic
connectivity algorithms mentioned in the introduction can also
maintain the number of connected components, so the same almost-tight
upper bounds hold for this problem.

We use the same graph as in the dynamic connectivity lower bound,
except that we add a new vertex $s$ which is connected to all nodes
from the first column. The updates in the connectivity problem
translate to identical updates in our current problem. The hard
instance of connectivity asks queries between a vertex $u$ on the
first column, and an arbitrary vertex $v$. To simulate these, we
disconnect $u$ from $s$, connect $v$ to $s$, and ask whether the
entire graph is connected; after this, we undo the two changes to the
graph. If $u$ and $v$ were on distinct paths, $u$'s path will now be
disconnected from the rest of the graph. Otherwise, the edge $(v,s)$
will reconnect the path to the rest of the graph.

The graph we consider is a tree, so it is plane regardless of the
embedding of the vertices. During a query, if $u$ and $v$ are on the
same path, we create an ephemeral cycle. However, the $(v,s)$ edge can
simply be routed along the old path $s \to u \leadsto v$, so the graph
remains plane.

\subsection{Dynamic MSF}
The problem is to maintain the cost of the minimum spanning forest in
a weighted dynamic graph. The problem can be solved in $O(\lg^4 n)$
time per operation \cite{holm01connect}. In plane graphs, the problem
admits a solution in time $O(\lg n)$ \cite{eppstein92planar}. We
obtain a lower bound of $\Omega(\lg n)$, which holds even for plane
graphs with unit weights. Our bound follows immediately from the
previous bound. If all edges have unit weight and the graph is
connected, the weight of the MSF is $n-1$. If the graph is
disconnected, the weight of the MSF is strictly smaller.

\subsection{Dynamic Planarity Testing}
The problem is to maintain a dynamic plane graph, and test whether
inserting a given edge would destroy planarity. Actual insertions
always maintain planarity; an edge $(u,v)$ is given along with an
order inside the set of edges adjacent to $u$ and $v$. The problem can
be solved in $O(\lg^2 n)$ time per operation
\cite{italiano93planarity}. A lower bound of $\Omega(\lg n / \lg\lg
n)$ appears in \cite{fredman98connect}. We obtain a lower bound of
$\Omega(\lg n)$.

Because the graph from our lower bound proof is always a collection of
disjoint paths, it is plane under any embedding. Consider on the side
the complete bipartite graph $K_{3,3}$, from which an edge $(s,t)$ is
removed. Without that edge, this annex graph is also planar. To
implement connectivity queries between two nodes $u$ and $v$, we first
insert the edge $(u,s)$ temporarily, and then query whether inserting
the edge $(v,t)$ would destroy planarity. If $u$ and $v$ are on
distinct paths, the graph created by adding $(u,s)$ and $(v,t)$ is
planar, and can be embedded for any relative order of these two edges
(the edges of $K_{3,3} \setminus \{(s,t)\}$ can simply go around the
two paths containing $u$ and $v$). If $u$ and $v$ are on the same
path, we would be creating a subdivision (graph expansion) of
$K_{3,3}$, so the graph would no longer be planar (by Kuratowski's
theorem).

\section{Open Problems} \label{theend}

This paper provides powerful techniques for understanding problems
which have complexity around $\Theta(\lg n)$. The chronogram technique
had already proven effective for problems with complexity
$\Theta(\frac{\lg n}{\lg\lg n})$. However, current techniques seem
helpless either below or above these thresholds. Below this regime, we
have integer search problems, such as priority queues. Looking at
higher complexities, we find many important problems which seem to
have polylogarithmic complexity (such as range queries in higher
dimensions) or even $n^{\Omega(1)}$ complexity (dynamic problems on
directed graphs). It is also an important complexity theoretic
challenge to obtain an $\omega(\lg n)$ lower bound for a dynamic
language membership problem.

It is also worth noting that our bounds do not bring a complete
understanding of the partial-sums problem when $\delta = o(b)$.
First, we cannot prove a tight bound for \func{verify-sum}. A bound of
$\Omega(\lg n / \lg b)$, for any $\delta$, is implicit in
\cite{husfeldt03sums}, and can also be reproved using our
technique. Second, we do not have a good understanding of the possible
trade-offs. For \func{select}, this seems a thorny issue, because of the
interaction with the predecessor problem. Even for \func{sum}, we do
not know what bounds are possible in the range $t_u < t_q$. It is
tempting to think that the right bound is $t_u (\lg \frac{t_q}{t_u} +
\lg \frac{b}{\delta}) = \Theta(\lg n)$, by symmetry with the case $t_u
> t_q$. However, buffer trees \cite{arge03buffer} give better bounds
for some choices of parameters, e.g.~when $b = \Omega(\lg^{1+\eps}
n)$. This problem seems to touch on a fundamental issue: a good lower
bound apparently needs to argue that the data structure has to recover
a lot of information about the indices of updates, in addition to the
$\Delta$ values.

It would be very interesting to obtain a logarithmic upper bound for
dynamic connectivity, matching our lower bound. It would also be
interesting to determine the complexity of decremental connectivity.
For this problem, at least our trade-off lower bound cannot hold,
because \cite{henzinger99connect} gave a solution with
polylogarithmic updates and constant query time.

\bibliographystyle{alpha}
\bibliography{../general}

\end{document}